\documentclass[11pt,a4paper,twoside]{article}

\usepackage[utf8]{inputenc}
\usepackage[T1]{fontenc}

\usepackage{amsmath,amssymb,amsthm,amsfonts}
\usepackage{bm}           
\usepackage{mathtools}    
\usepackage[toc,page]{appendix}
\usepackage[makeroom]{cancel}
\everymath{\displaystyle}
\allowdisplaybreaks

\newtheorem{lemma}{Lemma}

\newtheorem{definition}{Definition}

\usepackage[dvipsnames]{xcolor}   
\usepackage{graphicx}             
\usepackage{rotating}             
\usepackage{array}                
\usepackage{float}                
\usepackage{subcaption}           
\usepackage[labelfont=bf,font=small]{caption}

\usepackage[dvipsnames]{xcolor}

\usepackage{tikz}
\usetikzlibrary{shapes.geometric, arrows}

\usepackage[margin=1in]{geometry}

\tikzstyle{process} = [rectangle, minimum width=3cm, minimum height=1cm, text centered, draw=blue!50, fill=blue!20, text width=3cm]

\usepackage{booktabs,threeparttable,arydshln,longtable,siunitx}
\usepackage{tabularx}
\usepackage{multirow,multicol,blkarray,array}
\usepackage[section]{placeins}  
\usepackage{numprint}

\usepackage{epigraph}
\usepackage{marginnote}
\usepackage{soul}         
\usepackage{lscape, pdflscape}
\usepackage{indentfirst}
\usepackage{enumitem}
\usepackage{cases}
\usepackage{url}
\usepackage{setspace}
\usepackage{ulem}         
\usepackage{framed}
\usepackage{adjustbox}
\usepackage{here}
\usepackage{listings}
\usepackage{xstring}
\usepackage{flafter}
\usepackage{comment}

\usepackage[
  colorlinks   = true,
  linkcolor    = red,
  citecolor    = blue,
  urlcolor     = cyan,
  filecolor    = black,
  bookmarks    = false,
  pdfauthor    = {Felix Kubler},
  pdfsubject   = {},
  pdftitle     = {}
]{hyperref}

\usepackage{cleveref}
\crefname{equation}{Eq.}{Eqs.}
\crefname{figure}  {Figure}{Figures}
\crefname{table}   {Table}{Tables}
\crefname{line}    {Algorithm}{Algorithms}
\crefname{asmp}    {Assumption}{Assumptions}
\crefname{section} {Section}{Sections}
\crefname{chapter} {Chapter}{Chapters}
\crefname{appsec}  {Appendix}{Appendixes}

\renewcommand{\ref}{\cref}  

\usepackage{natbib}

\usepackage{fancyhdr}
\fancypagestyle{firstpagestyle}{
  
  \fancyhf{}
  \fancyfoot[C]{\thepage}
}

\fancypagestyle{CVfirstpagestyle}{
  
  \fancyhf{}
  \fancyfoot[C]{\thepage}
}

\DeclareFontFamily{U}{mathx}{}
\DeclareFontShape{U}{mathx}{m}{n}{<-> mathx10}{}
\DeclareSymbolFont{mathx}{U}{mathx}{m}{n}
\DeclareMathAccent{\widehat}{0}{mathx}{"70}
\DeclareMathAccent{\widecheck}{0}{mathx}{"71}

\newcommand{\R}{\mathbb{R}}

\newcommand{\argmax}{\mathop{\rm arg\,max}\limits}

\newcommand{\expnum}[2]{\ifnum#1=1 10^{#2} \else #1\!\cdot\!10^{#2}\fi}

\newcommand{\Ky}{K_{\bm{y}}}
\newcommand{\bx}{\bm{x}}
\newcommand{\by}{\bm{y}}

\renewcommand{\emph}[1]{\textit{#1}}

\begin{document} 
\title{Bayesian Optimization on the Equilibrium Manifold\thanks{I thank
seminar participants at the PHBS UK Conference on the Frontiers of AI in Economics and in particular
  Tom Sargent and Simon Scheidegger for their comments. I used Claude Opus 4.8 for coding assistance. I thank the SNF for financial support.}}

\author{
    Felix Kübler\thanks{Department of Finance, University of Zurich, Switzerland; Email: \href{mailto:felix.kuebler@df.uzh.ch}{felix.kuebler@df.uzh.ch}}
}
\date{\today}

\maketitle

\begin{abstract}
Computing optimal policy in heterogeneous-agent economies is complicated by the possibility of multiple equilibria. We overcome this difficulty by showing that when the equilibrium manifold has a low-dimensional Negishi-weight parameterization, Bayesian optimization reliably finds approximate solutions and can be used to certify candidate solutions with high probability. This insight brings recent machine learning advances to bear on a core problem in macroeconomics.
We apply Bayesian optimization to a dynamic economy with heterogeneous agents and climate change and compute optimal carbon taxes in this setting. Although in principle the presence of the carbon externality creates scope for multiple equilibria, we show that in an example with realistic calibration of damages competitive equilibra are most likely unique.

\end{abstract}


\newpage

\section{Introduction}

Many interesting questions about optimal economic policy can be formulated as non-convex  optimization problems with equilibrium constraints -- prominent examples include
 models of optimal fiscal policy (\cite{aiyagari2002optimal}, \cite{dyrda2023optimal}), models of optimal carbon policy (\cite{nordhaus1996regional}, \cite{kotlikoff2021making}) or models of macro-prudential regulation (\cite{davila2018pecuniary}, \cite{jeanne2019managing}).
In general, there are only few results that justify the use of a first order approach to characterize optimal solutions in these models. Moreover, in the presence of heterogeneous agents, the possibility of multiple equilibria seems to make it impossible to reliably compute the optimal policy. In this paper, we consider an economically meaningful relexation of the original problem that allows us to write it as a box-constrained maximization problem, and we explore how Bayesian optimization (see \cite{garnett2023bayesian}) can be used to solve this problem efficiently. Our key insight is that reliable probabilistic solutions can be obtained with moderate computational effort in cases where deterministic solutions are not tractable.

To explain this paper's contribution, it is useful to consider the following abstract mathematical problem.
\begin{equation} \label{p1} \max_{x\in B \subset {\mathbb R}^n} f(x) \mbox{ s.t. } h_i(x) \le 0, i=1, \ldots, K. \end{equation}
Without convexity assumptions, this problem is generally not tractable and cannot be solved for moderate values for $n$.
However, in economic models,
it is often the case that one can write $ x=(x^1,x^2) \in B_1 \times B_2  $, where $ B_1 $ is a $ n_1 $ dimensional box and $ B_2 $ is a $ n_2 $ dimensional box with $ n_2 \gg n_1 $ and that 
there exists a function $ \tilde{x}^2:B_1 \rightarrow B_2 $ as well a function $ g: B_1 \rightarrow {\mathbb R}^L $ such that (\ref{p1}) is equivalent to
\begin{equation} \label{p2} \max_{x^1\in B_1 } f(x^1,\tilde{x}^2(x^1)) \mbox{ s.t. } g_i(x^1) \le 0, i=1, \ldots, L. \end{equation} 
Moreover, one can often define a family of penalty functions
$ ({\cal P}(\cdot,\lambda))_{\lambda \ge 0} $
such that the solution to the relaxed problem
\begin{equation} \label{p3} \max_{x^1 \in B_1} V(x^1)=f(x^1,\tilde{x}^2(x^1)) - {\cal P}(g(x^1),\lambda)  \end{equation}
has a sensible economic interpretation for any $ \lambda > 0 $ and such that the solution to (\ref{p3}) converges to the solution of (\ref{p2}) as $ \lambda \rightarrow \infty$.

So far, it is not clear whether (\ref{p3}) is easier to solve than (\ref{p1}). 
In most cases, the function $\tilde{x}^2(x^1)$ is only implicitly defined as the solution of a system of nonlinear equations. If this is the case, one first has to verify that there exists a unique solution. The computation of $\tilde{x}^2(x^1)$ becomes costly, and it might be impossible to compute derivatives. 

However, we assume that the decision maker is Bayesian and uses a computational method to determine an approximate solution to (\ref{p3}). For given $ \epsilon >0 , \delta $ the computation terminates if it produces a $ \hat{x}^1 $ that satisfies
\begin{equation}
\label{BPlanner}
\mbox{Prob}[\exists x^1: V(x^1)>V(\hat x^1)+\delta]< \epsilon,
\end{equation}
where the probability denotes the posterior probability given all the information inquired during the computations.
In this case, the  numerical methods developed for Bayesian optimization are ideally suited for solving (\ref{BPlanner}). 

Bayesian optimization is a strategy for finding the maximum of expensive-to-evaluate functions, and it is particularly useful when one does not have access to gradients. The problem of finding the maximum of an unknown function was first examined in \cite{hotelling1941experimental}.
\cite{friedmansavage} point out the advantages of a sequential design. In the modern treatment of the problem, based on \cite{Kushner},  
the core idea is to build a surrogate model of the objective function and use it to decide where to evaluate next (see \cite{garnett2023bayesian} for a detailed introduction to the topic).
This surrogate must be
a probabilistic model, such as a Gaussian process, as the
 probabilistic treatment of uncertainty is vital to global
optimization: one must be able to reason about
the uncertainty in unvisited regions of the objective function's domain. 

An important class of economic applications takes the inequality constraints in (\ref{p1}) to be equality constraints and conditions for competitive equilibrium, and $ x^1 $ to be a vector of policy instruments such as taxes and transfers. If one could show that equilibria are unique for all policies $ \tilde{x}^2(x^1)$ would simply be the map of policies to equilibrium prices and allocations. Unfortunately, in most models, the known sufficient conditions for uniqueness of equilibrium are fare to restrictive, even though multiplicity does not turn out to be a problem for most calibrations.
One direct implication of (\ref{BPlanner}) that often proves extremely useful in economic applications is that BO can be used to certify, with high probability, that a multivariate smooth function $f(x,\alpha)$ is monotone in $x$ for all parameters $ \alpha $. For this, one would take 
as the objective function in (\ref{BPlanner}) the largest eigenvalue of the symmetric part of the Jacobian of $ f $ with respect to $x$. In the context of (\ref{p1}),
monotonicity of $ h $ in $ x^2 $ is, of course, a sufficient condition for uniqueness. Therefore, one approach to solving (\ref{p1}) is often to establish that $ \tilde{x}^2(x^1) $ is  is in fact a function with high probability and to solve for it using Newton's method.

Unfortunately, monotonicity of $h$ often turns out to be too strong a condition for uniqueness. The strategy we propose in this paper is then to identify a subset of equations for which uniqueness of solution can be established, solve these equations with Newton's method, and include the remaining equations in the penalty term.

To illustrate these ideas, we first consider a general equilibrium model of an exchange economy, where we characterize competitive equilibira with the Negishi-approach
(see, e.g. \cite{kehoe1992characterizing}). The penalty function in (\ref{p3}) can be taken to be the sum of squared budget violations across households; in the context of a policy problem, it can be interpreted as the costs of lump-sum transfers. We use a simple example of multiple equilibria to illustrate how BO correctly identifies them. 
We then extend the model to include a public good and 
 a (Ramsey) planner who produces the public good by confiscating the inputs from the households and who maximizes a social welfare function\footnote{The model mainly serves illustrative purposes, the method developed can be applied to a wide variety of optimal policy problems.}. 
In this context, one might naively 
 take $ x^1$ as a vector that specifies economic policy (supply of the public good) and $ x^2$ as equilibrium prices and allocations for that policy.
 The difficulty is that multiple equilibria generally cannot be ruled out: for a fixed policy, several equilibria may coexist, each yielding a different value of the planner's objective. Naturally, one would assume that the planner can select the equilibrium that yields the highest value of the social welfare function — but, as noted above, there is no continuous function $ \tilde{x}^2(x^1) $. 
We instead let $ x^1 $ consist of the public good supply together with all agents' Negishi weights. Given production decisions, consumption allocations are then the solution to a convex optimization problem and can be determined easily and reliably. We provide simple examples constructed to exhibit multiple equilibria and show that the BO algorithm correctly identifies the one that yields the highest value of the objective function.

To illustrate that the method applies to economically interesting settings, we consider a macroeconomic model of climate change with infinitely lived agents, polluting firms, and climate damages to both total factor productivity and depreciation. We give an example showing that even in a single-agent version of the model (similar to \cite{barrage2024policies}), the presence of externalities can give rise to multiple competitive equilibria. We then turn to a model with four heterogeneous agents and show how BO can be used to solve for the optimal carbon tax and revenue-sharing rule.
We proceed in two steps. First, we use BO to certify that for our calibration and for given taxes and Negishi weights, the equilibrium exhibits a unique capital path with high probability. Second, using BO together with penalties for budget violations, we show that in the absence of carbon taxes the equilibrium is unique with probability very close to one.
We consider both a constant carbon tax and a carbon tax that is an arbitrary time-dependent function. In our calibrations, carbon tax revenue is large enough that the planner can achieve the same solution as under unrestricted lump-sum transfers. The welfare differences between a simple constant-tax scheme and more elaborate schemes turn out to be insignificant. In the  case with constant carbon taxes, it can be certified that equilibria are unique with high probability.

While there is a large literature on the computation of optimal policies in dynamic economic models, there are few papers that apply ideas from machine learning to the problem.
\cite{scheidegger2019machine} are the first to use Gaussian process regression for dynamic economic modeling. \cite{chen2026deep} demonstrate how the use of a surrogate model can lead to large efficiency gains in option pricing problems, but neither of these papers considers Bayesian optimization.
Ignoring the problem of multiple equilibria,
\cite{kubler2025using} are the first to use Bayesian optimization to characterize optimal policies. They analyze constrained optimal carbon taxes in a stochastic OLG model with climate change. In that paper the computation of equilibria for a given set of instruments is one of the main contributions.
The main contribution of this paper is to use Bayesian optimization to find constrained optimal policies taking the possibility of multiple equilibria into account. 
As mentioned above, the results of BO used in this article are all standard and can be found in \cite{garnett2023bayesian}.

In this paper, the successful demonstrations of our method operate in relatively low dimensions. In some of the examples there might be possible alternatives to BO - dense grids in one or two dimensions, multistart deterministic global optimization, homotopy/continuation in Negishi weights,  deterministic global methods using Lipschitz bounds, or stochastic global methods for final certification offer standard benchmarks.  
It is subject to future research to establish how BO compares to plausible alternatives on the examples in this paper. Because acquisition maximization is inherently non-convex, defining the exact threshold where BO overtakes existing computational approaches  would solidify the framework's comparative advantage. At this point, the results are more suggestive than definitive.

The remainder of the paper is organized as follows. Section 2 gives a brief review of Bayesian optimization and Gaussian process regressions. We explain how probabilistic results can be derived with this method. Section 3 provides simple examples to illustrate the method. Section 4
applies the method to optimal carbon taxation in a climate change macro model with heterogeneous agents. Section 5 concludes. The Appendix contains further theoretical details on Gaussian process modeling and a discussion on the maximization of acquistion functions.

\section{Bayesian Optimization}
All the results in this section are standard and are repeated here only for completeness.
Good references are \cite{williams2006gaussian}, \cite{garnett2023bayesian} and \cite{hennig2022probabilistic}.

Bayesian Optimization (BO) is a strategy for global optimization of black-box functions that are expensive to evaluate. 
The key idea is to build a probabilistic surrogate model of the objective function and use it to decide where to sample next, balancing exploration and exploitation. 
Consider the global optimization problem:
\begin{equation}
\label{BO-formulation}
x^* = \arg\max_{x \in \mathcal{X}} f(x)
\end{equation}
where:
\begin{itemize}
\item $f: \mathcal{X} \to \mathbb{R}$ is the objective function (black-box, expensive to evaluate)
\item $\mathcal{X} \subseteq \mathbb{R}^d$ is a compact domain (often a hyperrectangle)
\item We can only observe $y_i = f(x_i) + \epsilon_i$ where $\epsilon_i \sim \mathcal{N}(0, \sigma_{\epsilon}^2)$ is i.i.d. noise.
\end{itemize}

The goal is to find $x^*$ (or a good approximation) using as few evaluations of $f$ as possible.
While in our experiment below, we typically consider the noise-free case with $ \sigma_{\epsilon}^2=0 $, it is standard to introduce the general concepts for the noisy case.
Bayesian optimization consists of two main components:
\begin{enumerate}
\item Surrogate Model:\\
A probabilistic model of the objective function, typically a \textbf{Gaussian Process} (GP):
\begin{equation}
f(x) \sim \mathcal{GP}(m(x), k(x, x'))
\end{equation}

\item Acquisition Function:\\
A function $\alpha: \mathcal{X} \to \mathbb{R}$ that uses the surrogate model to determine the next point to evaluate:
\begin{equation}
x_{n+1} = \arg\max_{x \in \mathcal{X}} \alpha(x \mid \mathcal{D}_n)
\end{equation}

The acquisition function balances exploitation (sample where the model predicts high values) and exploration (sample where uncertainty is high).
\end{enumerate}

In order to provide theoretical foundations for  the use of BO in economics, we first review the theory of GP's and then summarize known theoretical properties of BO algorithms. In particular, we explain how probabilistic results can be efficiently certified.
\subsection{Introduction to Gaussian processes}
\begin{definition}[Gaussian Process]
 A \textbf{Gaussian process} (GP) is a collection of random variables, any finite number of which have a joint Gaussian distribution. Formally, a stochastic process $\{f(x) : x \in \mathcal{X}\}$ is a Gaussian process if for any finite subset $\{x_1, \ldots, x_n\} \subset \mathcal{X}$, the random vector $(f(x_1), \ldots, f(x_n))^\top$ follows a multivariate normal distribution.
\end{definition}
It is easy to see that a Gaussian process is completely specified by its mean function and covariance function (kernel):
\begin{align}
m(x) &= \mathbb{E}[f(x)] \\
k(x, x') &= \mathbb{E}[(f(x) - m(x))(f(x') - m(x'))]
\end{align}
We write $ f \sim \mathcal{GP}(m, k)$. The general definition of a covariance kernel is as follows.
\begin{definition}[Covariance Kernel]
A function $k: \mathcal{X} \times \mathcal{X} \to \mathbb{R}$ is a \textbf{covariance kernel} (or positive definite kernel) if for any finite set of points $\{x_1, \ldots, x_n\} \subset \mathcal{X}$, the matrix $K$ with entries $K_{ij} = k(x_i, x_j)$ is positive semidefinite.
\end{definition}
Throughout this paper, we focus on the so-called (ARD) squared-exponential kernel defined as
$$ k_{SE}(x,x')=\sigma^2_f exp\left(-\frac{1}{2}\sum_{d=1}^D\frac{(x_d-x'_d)^2}{\ell_d^2} \right) .$$
ARD ("automatic relevance determination")  means that we specify  one lengthscale $ \ell_d $  per input dimension instead of a single shared lengthscale as in the case of the isotropic SE kernel.

The hyperparameters $ \sigma_f, (\ell_1,\ldots,\ell_d) $ play an important role and their determination is discussed in the appendix.
In the appendix we also introduce a more general parametric family of kernels, the Mat\'ern kernels, and discuss their properties.

\subsubsection{GP regression}
Consider the regression problem with training data $\mathcal{D} = \{(\mathbf{x}_i, y_i)\}_{i=1}^n$, where $\mathbf{x}_i \in \mathbb{R}^D$ are inputs and $y_i \in \mathbb{R}$ are noisy observations:
\begin{equation}
y_i = f(\mathbf{x}_i) + \epsilon_i, \quad \epsilon_i \sim \mathcal{N}(0, \sigma_{\epsilon}^2)
\end{equation}

We place a GP prior on $f$:
\begin{equation}
f \sim \mathcal{GP}(0, k(\mathbf{x}, \mathbf{x}'))
\end{equation}

Let $\mathbf{X} = [\mathbf{x}_1, \ldots, \mathbf{x}_n]^\top$ be the training inputs, $\mathbf{y} = [y_1, \ldots, y_n]^\top$ the training outputs, and $\mathbf{X}_* = [\mathbf{x}_{*1}, \ldots, \mathbf{x}_{*m}]^\top$ the test inputs. Define:
\begin{align}
K &= k(\mathbf{X}, \mathbf{X}) \in \mathbb{R}^{n \times n} \\
K_* &= k(\mathbf{X}_*, \mathbf{X}) \in \mathbb{R}^{m \times n} \\
K_{**} &= k(\mathbf{X}_*, \mathbf{X}_*) \in \mathbb{R}^{m \times m}
\end{align}

The joint distribution of observed and test function values is:
\begin{equation}
\begin{bmatrix} \mathbf{y} \\ \mathbf{f}_* \end{bmatrix} \sim \mathcal{N}\left(\mathbf{0}, \begin{bmatrix} K + \sigma_{\epsilon}^2 I & K_*^\top \\ K_* & K_{**} \end{bmatrix}\right)
\end{equation}

The posterior distribution of the test outputs $\mathbf{f}_*$ given the training data is Gaussian:
\begin{equation}
\mathbf{f}_* | \mathbf{X}, \mathbf{y}, \mathbf{X}_* \sim \mathcal{N}(\boldsymbol{\mu}_*, \boldsymbol{\Sigma}_*)
\end{equation}
where
\begin{align}
\boldsymbol{\mu}_* &= K_*(K + \sigma_{\epsilon}^2 I)^{-1}\mathbf{y} \label{eq:mean}\\
\boldsymbol{\Sigma}_* &= K_{**} - K_*(K + \sigma_{\epsilon}^2 I)^{-1}K_*^\top \label{eq:cov}
\end{align}
For a single test point $\mathbf{x}_*$, the predictive mean and variance are:
\begin{align}
\mu(\mathbf{x}_*) &= \mathbf{k}_*^\top (K + \sigma_{\epsilon}^2 I)^{-1} \mathbf{y} \\
\sigma^2(\mathbf{x}_*) &= k(\mathbf{x}_*, \mathbf{x}_*) - \mathbf{k}_*^\top (K + \sigma_{\epsilon}^2 I)^{-1} \mathbf{k}_*
\end{align}
where $\mathbf{k}_* = k(\mathbf{X}, \mathbf{x}_*) \in \mathbb{R}^n$.
We start with a prior mean and covariance function for the Gaussian process.

\subsection{Bayesian optimization}
Throughout the paper, we assume that the true objective function $ f(.) $ in (\ref{BO-formulation}) is a sample path of the Gaussian process used to model it. In a Bayesian analysis, this is the natural assumption. It provides a known distribution for the
function values at unobserved locations and therefore the foundation of any acquisition function.
 We discuss the assumption in Appendix \ref{rkhs}.

\subsubsection{Acquisition Functions for BO}
While there are a variety of different approaches to derive useful acquisition function (see \cite{garnett2023bayesian} for details) this paper focuses on the following two.
\begin{itemize}
\item Probability of Improvement (PI):
Let $f^+ = \max_{i=1,\ldots,n} f(x_i)$ be the best observed value. The probability of improvement is:
\begin{equation}
\text{PI}(x;\xi) = \mbox{Prob}(f(x) \geq f^+ + \xi) = \Phi\left(\frac{\mu_n(x) - f^+ - \xi}{\sigma_n(x)}\right)
\end{equation}
where $\Phi$ is the standard normal CDF and $\xi \geq 0$ is a trade-off parameter.
\item Upper Confidence bound (UCB):
Let $\beta_n > 0$ be a scalar exploration parameter. The \emph{upper confidence
bound} (UCB) acquisition function at iteration $n$ is defined by
\[
  \alpha_n^{\mathrm{UCB}}(x) \;:=\; \mu_n(x) \;+\; \sqrt{\beta_n}\,\sigma_n(x),
  \qquad x \in {\mathcal X}.
\]
The UCB acquisition function can be interpreted as a deterministic upper confidence
bound on $f(x)$: under the GP model,
\[
  \mbox{Prob}\!\left(f(x) \leq \mu_n(x) + \sqrt{\beta_n}\,\sigma_n(x)\right)
  \;=\; \Phi\!\left(\sqrt{\beta_n}\right),
\]
where $\Phi$ is the standard normal CDF. Choosing $\beta_n$ thus corresponds to
specifying a pointwise confidence level for the upper bound.
\end{itemize}

\subsubsection{Regret bounds and certification of global optima}
\label{sec:regret-bounds}
 Let $ x^* $ denote the solution to (\ref{BO-formulation}) and define
 the \emph{simple regret} after $T$ rounds:
\[
  R_T^{\mathrm{simple}} \;=\; f(x^*) - f(\hat{x}), \mbox{ where } f(\hat x)=\max_{i \le T} f(x_i).
\]

 In general we cannot obtain useful deterministic bounds on the simple regret, but we demonstrate in this section how to derive a probabilistic bound, i.e. for small $ \delta, \epsilon> 0 $ we want to make a statment of the form
\begin{equation} \label{goodbound} \mbox{Prob}\left[R_T^{\mathrm{simple}}\ge \delta \right] < \epsilon. 
\end{equation}
Using a PI acquisition function, with a trade-off parameter $ \delta $ and some small $ \eta>0 $, the algorithm can 
obviously be terminated after $ T $ rounds if $ PI(x) < \eta $ for all $ x \in \mathcal{X} $. However, whether this information suffices to obtain a bound as in (\ref{goodbound}) clearly depends on the Lipschitz continuity of the true objective function, $f$. If little prior knowledge of the unknown function $f(\cdot)$ is given, it might not be possible to directly derive a Lipschitz constant $L_f$ on ${\mathcal X}$. However, we indirectly assume a certain distribution of the derivatives of $f(\cdot)$ with the assumption that the true objective is a sample path of our GP. Therefore, it is possible to derive a probabilistic Lipschitz constant $L_f$ from this assumption, which is described in the following lemma from \cite{lederer2019uniform}.

\begin{lemma} Consider a zero mean Gaussian process defined through the covariance kernel $k(\cdot, \cdot)$ with continuous partial derivatives up to the fourth order and partial derivative kernels
$$
k^{\partial i}\left(\boldsymbol{x}, \boldsymbol{x}^{\prime}\right)=\frac{\partial^2}{\partial x_i \partial x_i^{\prime}} k\left(\boldsymbol{x}, \boldsymbol{x}^{\prime}\right) \quad \forall i=1, \ldots, D
$$

Let $L_k^{\partial i}$ denote the Lipschitz constants of the partial derivative kernels $k^{\partial i}(\cdot, \cdot)$ on the set $\mathcal{X}$ with maximal extension $r=\max _{\boldsymbol{x}, \boldsymbol{x} \prime \in \mathcal{X}}\left\|\boldsymbol{x}-\boldsymbol{x}^{\prime}\right\|$. Then, a sample function $f(\cdot)$ of the Gaussian process is almost surely continuous on $\mathcal{X}$ and with probability of at least $1-\delta_L$, it holds that
$$
L_f=\left\|\left[\begin{array}{c}
\sqrt{2 \log \left(\frac{2 D}{\delta_L}\right)} \max _{\boldsymbol{x} \in \mathcal{X}} \sqrt{k^{\partial 1}(\boldsymbol{x}, \boldsymbol{x})}+12 \sqrt{6 D} \max \left\{\max _{\boldsymbol{x} \in \mathcal{X}} \sqrt{k^{\partial 1}(\boldsymbol{x}, \boldsymbol{x})}, \sqrt{r L_k^{\partial 1}}\right\} \\
\vdots \\
\sqrt{2 \log \left(\frac{2 D}{\delta_L}\right)} \max _{\boldsymbol{x} \in \mathcal{X}} \sqrt{k^{\partial d}(\boldsymbol{x}, \boldsymbol{x})}+12 \sqrt{6 D} \max \left\{\max _{\boldsymbol{x} \in \mathcal{X}} \sqrt{k^{\partial D}(\boldsymbol{x}, \boldsymbol{x})}, \sqrt{r L_k^{\partial D}}\right\}
\end{array}\right]\right\|
$$
is a Lipschitz constant of $f(\cdot)$ on X.
\end{lemma}

For the SE kernel, it is straightforward to show that 
$$
\max _{\boldsymbol{x} \in \mathcal{X}} \sqrt{k(\boldsymbol{x}, \boldsymbol{x})}=\sigma_f \quad \max _{\boldsymbol{x} \in \mathcal{X}} \sqrt{k^{\partial i}(\boldsymbol{x}, \boldsymbol{x})}=\frac{\sigma_f}{\ell_i} .
$$

Furthermore, the Lipschitz constant of the derivative kernels is given by
$$
 L_k^{\partial i}=\omega \frac{\sigma_f^2}{\ell_i^3},
$$
where
$
\omega=\sqrt{6(3-\sqrt{6})} \exp \left(\sqrt{\frac{3}{2}}-\frac{3}{2}\right)
$.
Therefore, for the special case of a SE kernel, the 
 the bound for each dimension $i$ simplifies to:
  \begin{equation} 
  L_f^i = \frac{\sigma_f}{\ell_i}\sqrt{2\log\frac{2D}{\delta}} + 12\sqrt{6D}\cdot\max\left(\frac{\sigma_f}{\ell_i},   
  \sqrt{\frac{3 r\sigma_f^2}{\ell_i^3}}\right).\end{equation}

From this result, we can use a standard covering argument to obtain probabilistic bounds on simple regret. In the following, we do not carry through the probability of the Lipschitz bound $\delta_L$, it is understood that this needs to be factored into all the probabilistic statements that follow.

To obtain a bound of the form (\ref{goodbound}), we assume, for simplicity, that $ {\cal X} $ is a hyperrectangle. 
For some small $ \eta>0 $, we build a rectangular grid with spacing
  $ h_d = \frac{2\eta}{ D L_f^d} $
along each dimension $d$, resulting in $n_{\mathrm{cover}}$ points altogether. In what follows we refer to this as the covering number.

So for any $x \in \mathcal{X}$ its nearest
grid point $x_g$ satisfies:
\[
  f(x) - f(x_g)
  \;\leq\; \sum_d L_f^d \,\lvert x_d - x_{g,d}\rvert
  \;\leq\; \sum_d L_f^d \,\frac{h_d}{2}
  \;=\; \eta.
\]
 
So if $f(x) \geq \delta $, then $f(x_g) \geq \delta - \eta$ must hold.   Using the Lipschitz covering as well as a standard union bound, we obtain 
\begin{align*}
  \mbox{Prob}\left(\exists\, x \in \mathcal{X} : f(x) \geq \delta\right)
  &\;\leq\; \mbox{Prob}\left(\exists \text{ grid point } x_g : f(x_g) \geq \delta - \eta\right) \\
  &\;\leq\; \sum_{x_g} \mbox{Prob}\left(f(x_g) \geq \delta - \eta\right) \\
  &\;\leq\; n_{\mathrm{cover}} \cdot \max_{x_g}\,\mbox{Prob}\left(f(x_g) \geq \delta - \eta\right).
\end{align*}
 
With the definition of the PI acquisition function, we obtain 
\begin{align*}
  \mbox{Prob}\left(\exists\, x \in \mathcal{X} : f(x) \geq f(\hat x) + \gamma\right)
  &\;\leq\;
  n_{\mathrm{cover}}
  \cdot
  \max_{x \in \mathcal{X}}\;
  \mathrm{PI}\!\left(x;\; \gamma - \sum_d L_f^d \,\frac{h_d}{2}\right).
\end{align*}

By the same logic, we can partition $ {\mathcal X}$ into a $ \epsilon-$neighborhood around the observed current best point $ \hat{x}$ that satisfies $ f(\hat{x})=\max_{t \le T} f(x_t)$, $ {\mathbf B}_{\epsilon}(\hat x)$ and its complement in $ {\cal X} \setminus {\mathbf B}_{\epsilon}(\hat x)$. We then obtain bounds of the form
\begin{equation}
    \label{bound2}
 \mbox{Prob}\left(\exists x \in {\mathcal X} \setminus {\mathbf B}_{\epsilon}(\hat x): f(x)\ge f(\hat x)-\delta \right) < \eta.  \end{equation}

 Depending on the economic problem the bound (\ref{bound2}) might be more relevant than (\ref{goodbound}) -- it rules out that there are other local optima close-by and hence can ensure that equilibria are unique not only at given parameter-values but also in a neighborhood of these values. We will illustrate the usefulness of both bounds below. 
\subsection{Probably monotone functions}
\label{probmonfun}
A function $ f : {\mathbb R}^n \rightarrow {\mathbb R}^n $ is strictly monotonically decreasing whenever $ D_x f(x) $ is negative definite. In the context of multiplicity of equilibrium, monotonicity is an interesting property because montonone demand (or monotone expenditure) is a simply sufficient condition for uniqueness. To use BO to verify monotonicity, recall that a symmetric matrix is negative definite if and only if all of its eigenvalues are negative. One possible approach is to use BO to maximize the largest eigenvalue of $ D_xf(x)+(D_xf(x))^T$. Unfortunately, this is not a smooth function in $x$. We therefore consider the following objective function.
\begin{equation} \label{fun:mon} g(x)=\frac{1}{\beta} \log \left(\sum_{i=1}^n \exp(\beta \lambda_i(D_xf(x)+(D_xf(x))^T))\right), \end{equation}
where $ \lambda(M) $ denotes the vector of eigenvalues of a symetric matrix $M$.
It is easy to verify that
$$ \max_i \lambda_i(D_xf(x)+(D_xf(x))^T) \le g(x) \le 
\max_i \lambda_i(D_xf(x)+(D_xf(x))^T) + \log(n)/\beta. $$
It is also easy to verify that $ g(.) $ is a smooth function -- assuming that it is a sample path of a GP, we can use the analysis from the previous subsection to establish for small $ \delta, \epsilon > 0 $ that
$$ \mbox{Prob}(\forall x: g(x) \le -\delta) \ge 1-\epsilon .$$
We interpret this statement as saying that the function $f$ is probably monotone. We give examples of probably monotone functions in sections \ref{sec:probmonex} and \ref{sec:promoncap} below.

\subsection{Practical Considerations}
 
Gaussian process regression and Bayesian optimization have attracted a rich ecosystem of software
implementations. All computations reported in this paper have been performed using
\textbf{BoTorch} (see \cite{balandat2020botorch}). This is the current
state-of-the-art Python library, developed by Meta and built on
PyTorch (\cite{PyTorch}). Its key features are (i) auto-differentiation through
acquisition functions, enabling gradient-based optimization of the
inner loop; (ii) native GPU support; and (iii) a composable design that
makes it straightforward to implement custom models and acquisitions.

The examples in this paper are largely illustrative and solved without using GPU's. For higher dimensional problems, 
the Gaussian process (GP) regression requires forming and factorizing an
$n \times n$ kernel matrix $K$, whose $(i,j)$ entry is $k(x_i, x_j)$
for a chosen covariance function $k$. Cholesky factorization
$K = LL^\top$ costs $O(n^3)$ floating-point operations and $O(n^2)$
memory. This dominates the computational budget for moderate to large
$n$, and consists entirely of dense matrix operations -- precisely the
workload for which GPUs are optimized.
Moreover,
constructing the kernel matrix requires evaluating $k(x_i, x_j)$ for
all $\binom{n}{2}$ pairs. These evaluations are embarrassingly parallel
and memory-bandwidth-bound i.e. ideally suited for GPUs.

\section{Simple economic examples}
We first consider very simple examples which are known to have multiple equilibria and illustrate how BO correctly identifies multiplicity. We then introduce a public good and show how BO can be used to compute the optimal supply of this public good in the presence of multiplicty.
\subsection{Pure exchange economies}
We consider a pure exchange  economy with $H$ agents, $L$ private commodities. Each agent has a CES-utility function over consumption and endowments  $ \omega^h \in {\mathbb R}^L_+$.
For reasons that become clear below, it will be useful to characterize competitive equilibria via the Negishi approach. For this, define
$$ X(C,\lambda) = (X_h(C,\lambda))= \arg\max_{(c^h) \in {\mathbb R}^{HL}_+} \sum_{h\in {\mathbf H}} \lambda_h u^h(c)  \mbox{ s.t. } \sum_{h \in {\mathbf h}} c^h =C, $$
and 
$$ P(C,\lambda) = \frac{1}{D_{x_1} u_1 (X_1(C,\lambda))}D_x u_1 (X_1(C,\lambda)) .$$
A competitive equilibria is characterized by
$$ f(\lambda)=- (P(\omega,\lambda) \cdot (X^1(\omega,\lambda)-\omega^1), \ldots, P(\omega,\lambda) \cdot (X^H(\omega,\lambda)-\omega^H)) = 0 $$

\subsubsection{Identifying multiple equilibria}
We use variations of the example of multiple equilibria from
\cite{kehoe1991computation}
to illustrate how BO can correctly identify the multiplicity.
For this, suppose that $H=L=2$, utilities are
$$ u_1(x_1,x_2)=\frac{1024}{1-\gamma} x_1^{1-\gamma} + \frac{1}{1-\gamma} x_2^{1-\gamma}, \quad u_2(x_1,x_2)=\frac{1}{1-\gamma}x_1^{1-\gamma} +\frac{1024}{1-\gamma} x_2^{1-\gamma},$$
and endowments are
$ \omega^1=(12,1), \quad \omega^2=(1,12) $.

To illustrate basic principles of GP regression and BO we  first examine the following function of one variable. 
$$ \tilde{f}(\lambda)=- \left( P((\omega_1, \omega_2),\lambda) \cdot (X^1((\omega_1 , \omega_2),\lambda)- (\omega^1_1, \omega^1_2)') \right)^2 , \lambda\in [0,1] .$$ 
A competitive equilibrium can be characterized by $ \tilde{f}(\lambda)=0$ and all competitive equilibria can be found by identifying all maxima of the function $ \tilde{f}(.)$.
We show how this function can be maximized using BO.

We use BO with the UCB acquisition function and with $ \beta_n=3 $ for all $n$ to maximize $f$. We use 30 Sobol points for seeding.
We consider three different cases for $ \gamma $ - for $ \gamma=4$ the function $f$ is well behaved and has a unique local (and global) maximum. For $ \gamma=4.5$ there is still a unique local maximum, but the function becomes very flat around the maximum. For $ \gamma=5 $, there are 3 local maxima which all have the same value.

\begin{figure}[H]
    \centering
        \includegraphics[width=0.8\textwidth]{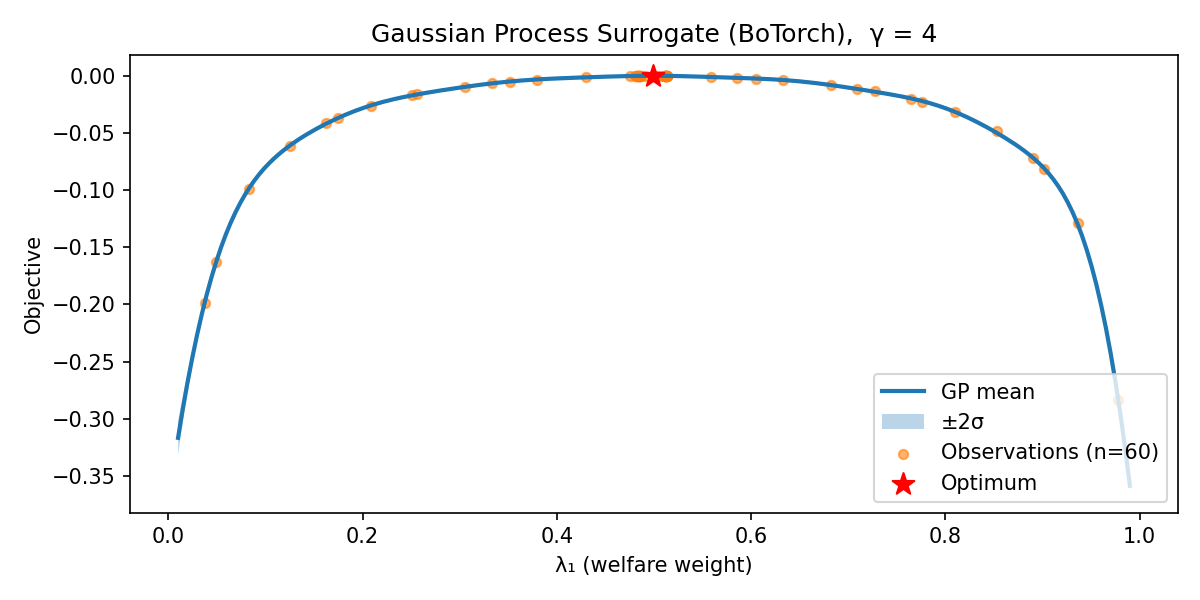}
        \hfill
        \includegraphics[width=0.8\textwidth]{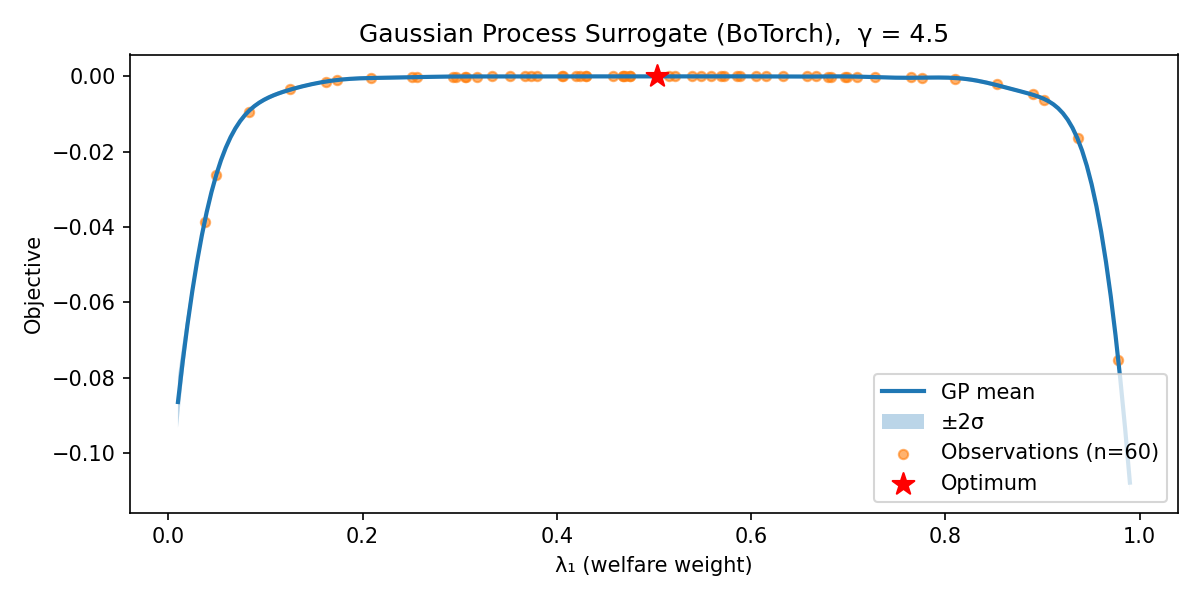}
                \hfill
        \includegraphics[width=0.8\textwidth]{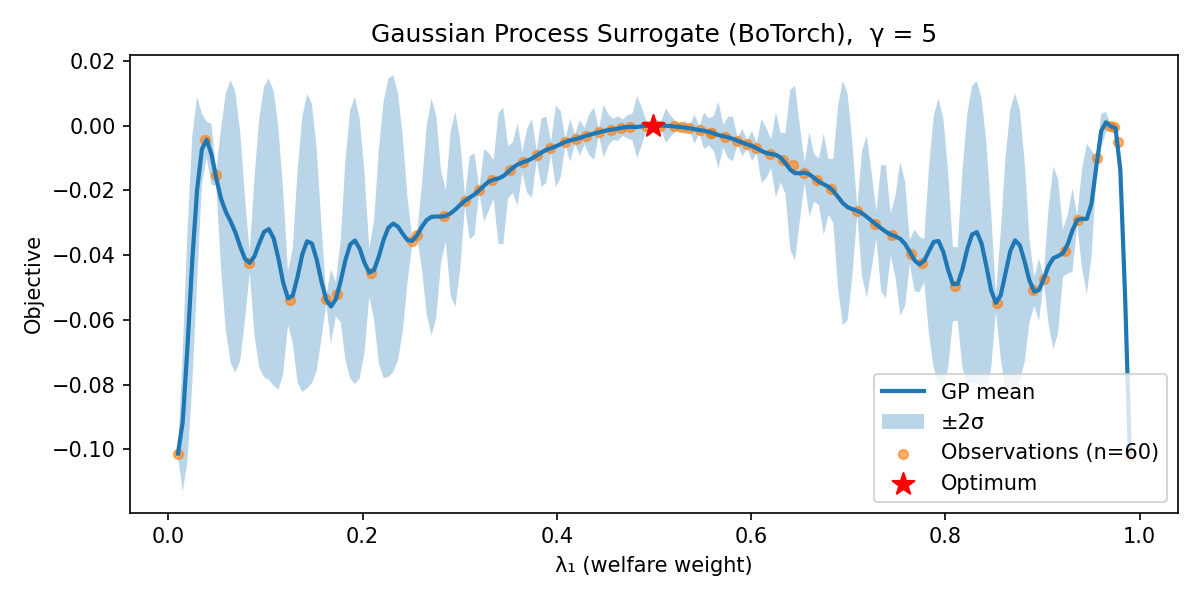}
                        \hfill
        \includegraphics[width=0.8\textwidth]{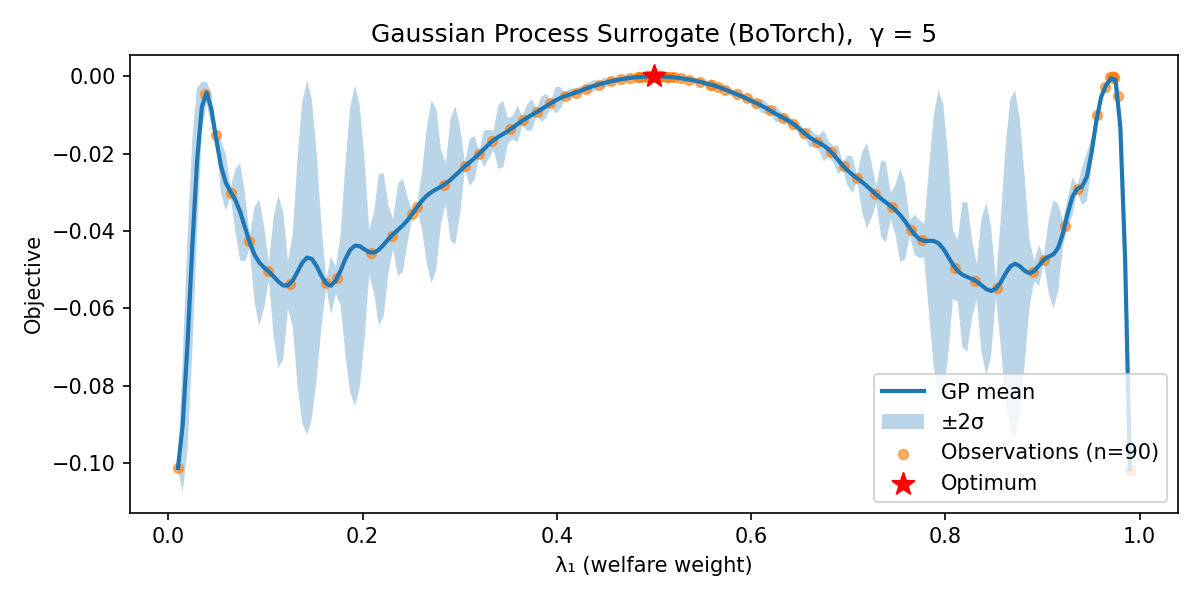}
                \caption{GP for different value of $ \gamma$}
    \label{fig:f1}
\end{figure}

Figure \ref{fig:f1} shows the Gaussian process (as mentioned above, we use a SE kernel and BoTorch for numerical implementation) after 30 BO iterations (together with the Sobol seed for a total of 60 points) for the three cases. 
For $ \gamma=4$ and for $ \gamma=4.5 $ the BO cleanly identifies the optimum, while the presence of multiple maxima makes it much harder to learn $ f(.) $ for $ \gamma=5 $ than for the two other cases.
However, if we increase the number of BO iterations from 30 to 60, the lowest graph in the figure shows that the three local optima can be identified with high probability. As can be seen from the lowest graph, there is no point where the probability that the function exceeds zero is greater than 2.5 percent (one-sided exceedence outside of two standard deviations of the standard normal distribution)\footnote{Note that this is a much weaker statement than (\ref{goodbound}), but still an interesting observation.}.

The figure shows that 
for $\gamma=5 $ the resulting pure exchange economy has three competitive equilibria which we can conveniently characterize by the associated Negishi weights $ \lambda_1^i $, i=1,2,3. It can be verified that $ \lambda_1^1 \simeq 0.0284 $
$ \lambda_1^2 =0.5 $, and $ \lambda^3_1 \simeq 1-0.0284 $. 

Clearly, it is impossible to establish Statement (\ref{bound2}) in this context for $ \gamma=5$. All three local optima yield the same value of the objective function. 
In contrast, for $ \gamma=4$ this statement can be established after only 40 draws of the function. The probability that there is another equilibrium outside of a small neighborhood of $ \lambda=1/2$ can be shown to be negligable.

The simple example illustrates how GP's and BO can be successfully used to identify multiple equilibria. Unfortunately, this comes at the costs of many having to generate many sample points. In high dimensional problems, this might quickly lead to infeasibility. BO is therefore not necessarily the best method to find all competitive equilibria in models where multiplicity is prevalent.  However, in many realistically calibrated economic models, one has the hunch that multiplicity \lq \lq is rare\rq\rq.
While this statement can generally not be established formally, it means that in applications the relevant functions are likely to be well behaved. Even if the dimensionality of the problem becomes large, BO can be used fruitfully. It will not find multiple equilibria because there are none. But one will be able to make verify this fact with high probability, therefore providing some formal justification for the hunch that uniqueness of equilibria is typical in these models.
The case $ \gamma=4.0 $ provides a clean example. There are clearly no theoretical results on uniqueness, but after only a few BO iterations one can say that with high probability equilibrium is unique. In fact, the reason why it is easy to establish uniqueness in this example is that the expenditure function $ f(\lambda) $ is monotone in $ \lambda$.
\subsubsection{Probably monotone expenditures}
\label{sec:probmonex}
As explained above, monotonicity of functions can be verified with BO. To illustrate this, consider the following extension of the simple example. We assume $ H=3, L=6 $ and take
$ \omega^h_l=12 $ for $ h=l $ and $ h=2l$, $ \omega^h_l=1 $ otherwise. We take utility to be CES with identical $ \gamma $ 
across agents and with coefficients $ a^h_l$ being
$  a^h_l=1024 $ for $ h=l $ and $ 2h=l$, $ a^h_l=1 $ otherwise. 
We maximize the function $ g(.)$ as defined in (\ref{fun:mon}) over $ \lambda_1,\ldots\lambda_{H-1} $,
$ \lambda_h \ge 0.01, $ $ \sum_{h=1}^{H-1} \lambda_h \le 0.99 $, as well as over $ \gamma \in [2,4] $.
After starting 200 Sobol points and 150 BO iterations, the best observed was at $ \lambda = [0.059, 0.787, 0.154],$ $ \gamma = 4 $ with $ g=-0.188$. Establishing Liptschitz bounds
gives bounds per dimension of 606.1,  872.2,  and 140.03 with probability $ 0.95 $. Taking $ \eta $ in the grid construction to be 0.01, we can establish that with overall probability (incl. the probability of the Lipschitz bounds) of 0.941 the underlying function is monotone, which trivially implies that equilibrium is unique for all $ \gamma \in [2,4] $ with probability 0.941.

\subsection{A simple planning problem}
Now suppose that in the two-good example, there is an additional public good. The government has access to a linear technology to produce the public good using commodity 1 as an input. A policy confiscates a fraction $ \xi $  of each agents' endowments of commodity 1 and produces $ \xi \omega_1 $ units of the public good. Equilibrium prices guarantee that each agents' utility maximizing consumptions clear the markets for the private good.

The government solves the following maximization problem.
\begin{eqnarray}
&&    \max_{(x^h),p,\xi} \sum_h \mu_h \left( u_h(x^h)+\frac{1}{2} \sqrt{\xi \omega_1} \right) \mbox{ s.t.}\\
&& x^h \in \argmax u^h(x) \mbox{ s.t. } p_1 x_1+p_2x_2 \le p_1 (1-\xi) \omega^h_1 + p_2 \omega^h_2 \label{eqe1}\\
&& \sum_{h} (x_1^h-(1-\xi) \omega_1^h)=0, \quad \sum_{h} (x_2^h-\omega_2^h)=0 \label{eqe2}
\end{eqnarray}

While there are several numerical methods to solve this simple example, it is worthy to note, however, that a first order approach (finding solutions to the first order KKT conditions) is generally not a viable method. Since this is not a convex optimization problem, there can be many solutions to the KKT conditions, and it is generally not tractable to compute all of them. Even if they can be written as polynomial functions and all solutions can be computed numerically, it becomes quickly untractable to do so (see, e.g. \cite{kubler2010tackling}).

The basic idea of this paper is to parameterize solutions to (\ref{eqe1})-(\ref{eqe2}) by some low-dimensional vector and to perform Bayesian optimization over $ \xi $ and the space of parameters. An obvious candidate for these parameters would be prices $ p_1, (1-p_1)$ - a one dimensional sufficient statistics for the solution. BO would then be a simple two dimensional problem. However, since in our main applications the number of commodities is much larger than the number of agents, we choose a different route.
Following \cite{kehoe1992characterizing} it is then useful to characterize equilbirium via a Negishi approach. Even if the welfare theorems do not hold, this approach often turns to be more efficient than a characterization via prices. Moreoever, we argue below that this approach has a cleaner economic interpretation.

We can then consider the following variation of  the planning problem.
\begin{eqnarray} \label{negform} && \max_{\lambda\in[0,1],1 \ge \xi \ge 0 } \sum_h \mu_h \left(u^h(X^h(\omega_1 (1-\xi), \omega_2,\lambda)  + \frac{1}{2}\sqrt{\xi \omega_1}\right) - \\
\notag
&& \eta \left( P(\omega_1 (1-\xi), \omega_2,\lambda) \cdot (X^1(\omega_1 (1-\xi), \omega_2,\lambda)- ((1-\xi) \omega^1_1, \omega^1_2)') \right)^2 .\end{eqnarray}
The penalty-parameter $ \eta > 0$ controls the transfers necessary to implement the solution as a competitive equilibrium. As  $ \eta \rightarrow \infty $ the budget constraints hold exactly and the solution to (\ref{negform}) converges to the solution of the original problem.

Besides the fact that in this formulation the number of commodities can be taken to be very large, this formulation is advantagous because one has a simple and economic meaningful interpretation of errors. One can imagine a economy where transfers are not impossible, but costly. While these costs are not modeled as real costs, they cause disutility in the planner's objective\footnote{If we were to use prices as the parameterization of equilibria, an approximate solution would imply approximate market clearing, which is infeasible and has no simple economic interpretation.}.

We now focus on the case $ \gamma=4.5 $  and show that the optimal supply of the public good crucially depends on the planner's weights $ \mu $. 
In fact, the solution changes discontinuously with $\mu$ - while for $ \xi=0 $ the competitive equilibrium is unique, for $ \xi>0 $ there can be several equilibria, and the welfare maximizing choice of $ \xi $ which depends on $ \mu $ determines one of these equilibria.

We use BO to compute the optimal public good supply for different values of $ \mu $. For each computation, we use 100 Sobol points as seeds and then employ 200 BO iterations. In each case the value of the best oberved point was above the GP mean plus 2 standard deviations on the entire domain.

\begin{figure}[th!]
    \centering
    \begin{subfigure}{0.45\textwidth}
        \includegraphics[width=0.95\textwidth]{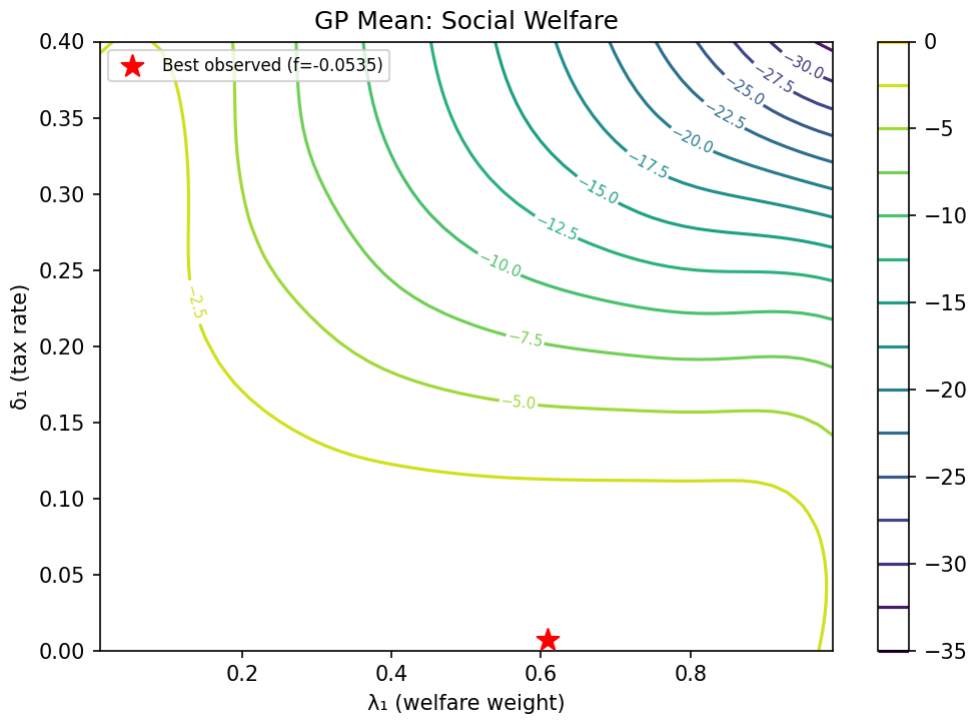}
        \caption{$\mu=(0.8,0.2)$}
        \label{fig:sub1}
    \end{subfigure}
    \hfill
    \begin{subfigure}{0.45\textwidth}
        \includegraphics[width=0.95\textwidth]{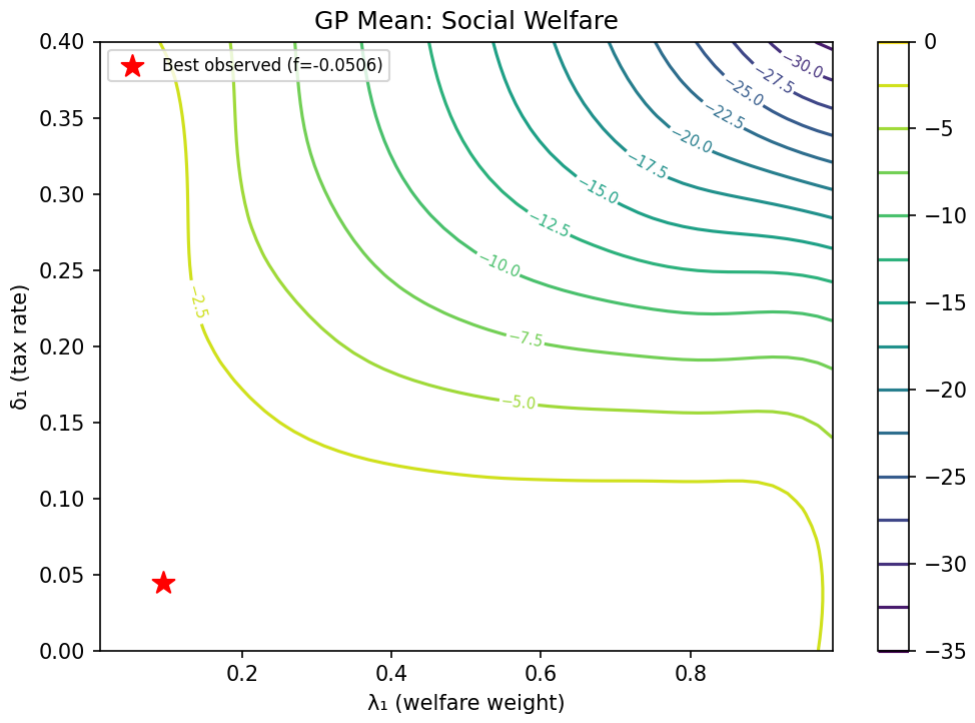}
        \caption{$ \mu=(0.7,0.3)$}
        \label{fig:sub2}
    \end{subfigure}

    \begin{subfigure}{0.45\textwidth}
        \includegraphics[width=0.95\textwidth]{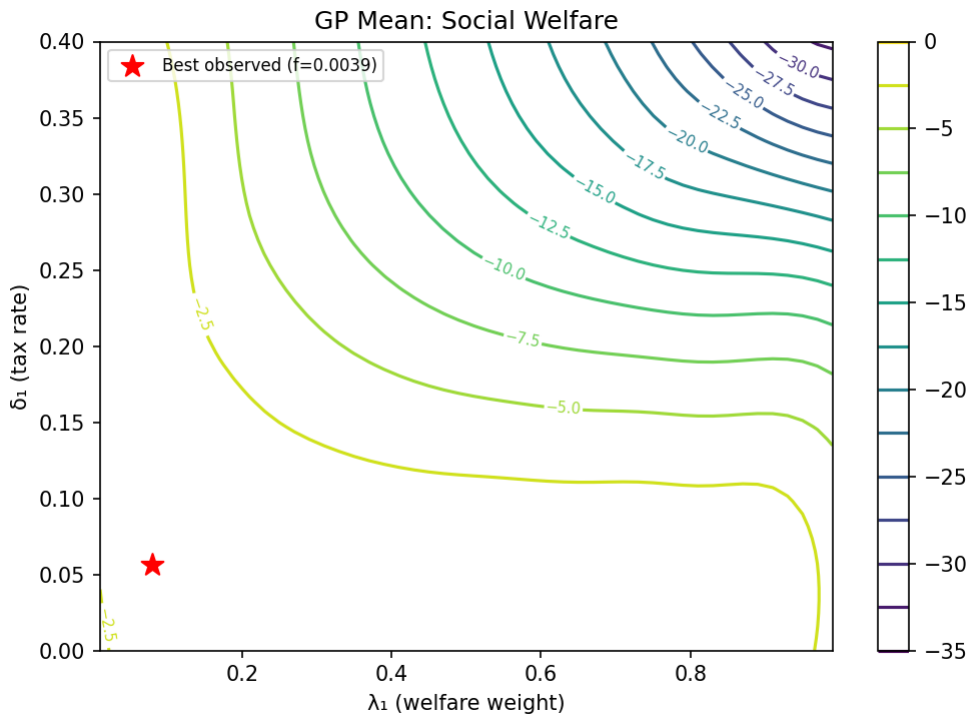}
        \caption{$\mu=(0.3,0.7)$}
        \label{fig:sub3}
    \end{subfigure}
    \hfill
    \begin{subfigure}{0.45\textwidth}
        \includegraphics[width=0.95\textwidth]{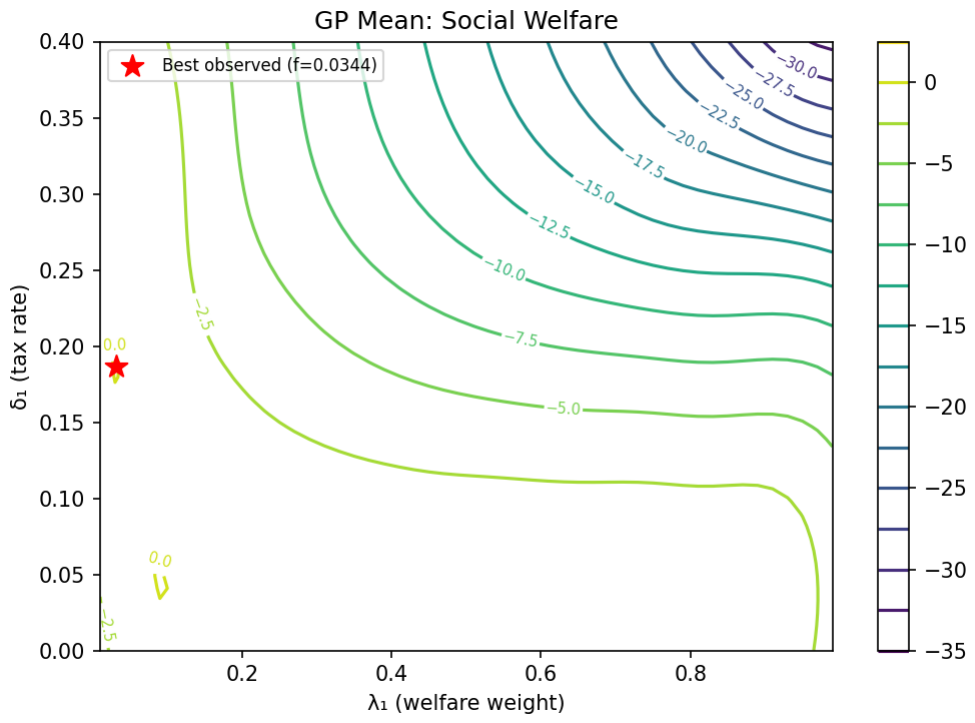}
        \caption{$\mu=(0.2,0.8)$}
        \label{fig:sub4}
    \end{subfigure}
    \caption{\textbf{Optimal Public Good Supply for Different Welfare Weights}}
    \label{fig:f2}
\end{figure}

Figure \ref{fig:f2} shows how the optimal solution crucially depends on the planner's weights. In fact, the solution jumps discontinuously as the weights change.
The optimal supply of the public good is very low if $ \mu $ is large. Agent 1 has a strong preference for good 1 and receives large disutiltiy from the production of the public good. Moreover, from the fact that there is less of good one available for consumption, its price becomes very high, decreasing consumption for agent 1. On the other hand, if agent 1 receives relatively little weight in the planner's objective, it is optimal to produce a large amount of the public good, pushing its price even higher and the utility of agent 1 to a very low level.

As can be clearly seen from the figure, the problem is not convex, and using first order conditions to obtain optimal solutions is likely to fail. In this simple example, it is not crucial to use BO for the computation. The example is meant to illustrate that BO can be used and that it efficiently finds the correct solution.

Next, we consider a more complicated model where, to the best of our knowledge, there are no other solution methods.

\section{Optimal carbon pricing}
As an application, we consider a dynamic model with  heterogeneous agents, capital accumulation, and a climate-externality.
The specification of production, abatement, and climate  change is from \cite{kubler2025using} which is a simplification of \cite{nordhausRevisitingSocialCost2017}.
As we show, an interesting aspect of the model is that even with a single agent, the presence of the externality implies that multiplicity of equilibria cannot be ruled out\footnote{This fact has received little attention in the literature. The possibility of multiple equilibri is typically only discussed in models with uncertainty and so-called tipping points, see, e.g.
\cite{lempert2006multiple}.}.
\subsection{The model}
Time is denoted by $ t=0,\ldots,T \le \infty.$ 
The representative firm produces the single consumption good $Y_t$ and the emissions $e_t$ at each time $t$:
\begin{equation}
(Y_t,e_t)
=
f_t(K_t,L_t,\mu_t)
=
\left(
\Phi(1-\mu_t) K_t^{\alpha} L_t^{1-\alpha}
+ (1 - \delta(T^{AT}_t)) K_t,\;
(1-\mu_t) \kappa_t K_t^{\alpha} L_t^{1-\alpha}
\right)
\end{equation}
Production is given by a Cobb-Douglas production function, where $\alpha$ {is the output elasticity of capital}, $K$, $L$, and $\delta$ are capital, labor, and depreciation, respectively. $ \mu $ denotes abatement and $ \Phi:[0,1] \rightarrow [0,1] $ is a mitigation-cost function. $\kappa$ is the exogenous carbon intensity of production, and $ T^{AT} $ denotes the global mean temperature anomaly above pre-industrial levels, which increases depreciation (and also decreases labor-supply). 
Following~\cite{nordhausRevisitingSocialCost2017}, we assume an abatement function of the form
\begin{equation}
\Phi (1- \mu) = 1 - \theta_{1} \mu^{\theta_2}.
\end{equation}
Note that when $e_t = 0$ (equivalently $\mu_t = 1$), output is reduced by a factor of $\theta_1$.

Firms take as given a carbon tax rate $\tau_t$ and maximize profits:
\begin{equation}
    \max_{\mu_t, K_t, L_t} (1-\theta_{1}\mu_t^{\theta_2}) K_t^{\alpha} L_t^{1-\alpha} + (1- \delta(T^{AT}_t) K_t -w_tL_t-(1+r_t)K_t- \tau_t (1-\mu_t) \kappa_t K_t^{\alpha} L_t^{1-\alpha}
    \label{eq:profit}
\end{equation}
$$\text{s.t.} \quad 0 \leq \mu_t \leq 1.$$

We assume that there are $H$ consumers that maximize time-separable utility over consumption and supply labor inelastically.
Agent $h$'s period utility is CRRA with risk aversion $\gamma_h$, and the time-discount
factor is constant at $\beta \in (0,1)$:
\[
  u_h(c) \;=\; \frac{c^{1-\gamma_h}}{1-\gamma_h}, \qquad
  U_h \;=\; \sum_{t=0}^{T} \beta^{t}\, u_h(c_t).
\]
At each time $t$, agent $h$'s labor-endowments are given by
$ \ell_t=\bar \ell_h(1-D_h(T^{AT}_t))$ for agent-specific damage function $ D_h$ . The government transfers carbon tax revenue to consumers, each agent $h$ obtains a fraction $ d_h \ge 0$ of the tax revenue $ \tau_t e_t $ in each period $t$.

Following ~\citet{BROCK2017263} and \cite{dietz2019cumulative}
we consider the simplest possible specification for clmate dynamics and assume that
\begin{equation}
    T_t^{AT} =\sigma_{CCR} \sum_{s=0}^{t-1} e_s , \sigma_{CCR}=1.7
\end{equation}
where $T_t^{AT}$ denotes the {global mean surface temperature anomaly above pre-industrial levels}, $e_s$ represents global emissions, and $ \sigma_{CCR}$ is the constant of proportionality, referred to as the carbon-climate response (CCR) or the Transient Climate Response to Cumulative Carbon Emissions (TCRE. \citet{matthews2009proportionality}) gives a scientific justification for this approximation, see \cite{Folini_2021} for a discussion of more realistic models of temperature dynamics.

The certainty-equivalent welfare is defined as
$u_h^{-1}\!\left(\bar U_h\right)$, where
$\bar U_h = U_h \big/ \sum_{t=0}^{T}\beta^{t}$. The social planner's objective is given by $$ W = \left(\sum_{h}  \left( u_h^{-1}\!\left(\bar U_h\right)\right)^{1-\gamma_P} \right)^{\frac{1}{1-\gamma_P}} ,$$  where $\gamma_p \ge 0 $ denotes his desire to smooth welfare across agents.

Initial capital is set to the (no-damage) steady-state level implied by
$\beta$ and $\delta_0$,
$$
  K_0 \;=\; \left(\frac{\alpha}{1/\beta - 1 + \delta_0}\right)^{1/(1-\alpha)},
$$
and each agent $h$ is assumed to hold an identical fraction of this initial capital.

\subsection{Competitive equilibrium for given taxes}
We first characterize competitive equilibria for a given sequence of taxes $ (\tau_t)_{t=0 \ldots T}$.
 From Equation (\ref{eq:profit}), we obtain that an interior optimal abatement choice $ \mu_t $ is determined uniquely by
 $ -\theta_1 \theta_2 \mu_t^{\theta_2-1} + \tau_t \kappa_t = 0 $.  The profit maximizing choices of $ K_t $ and $ L_t $ then imply
 $$ (1+r_t) = (1-\tau_t \kappa_t -\theta_1 \mu_t^{\theta_2}+\tau_t\mu_t \kappa_t) \alpha K_t^{\alpha-1} L_t^{1-\alpha}+(1-\delta(T^{AT}_t))$$
$$ w_t=  (1-\tau_t\kappa_t-\theta_1 \mu_t^{\theta_2}+\tau_t\mu_t \kappa_t) (1-\alpha) K_t^{\alpha} L_t^{-\alpha}. $$

It is easy to see that in  this model, for $ T<\infty $, given initial conditions $ K_0, \bar e_0 $, all equilibria can be parameterized by initial feasible individual consumption\footnote{Equivalently one can use aggregate consumption in period zero and Negishi-weights $ \lambda $ -- it will become clear below why it has advantages to use Negishi-weights.}
$ (c^h_0)_{h \in {\mathbf H}} \in {\mathbb R}^H_+$, $ \sum_h c^h_0<Y_0  $: Since labor is supplied inealistically, each $ L_t $ is determined by cumulative emissions at $ t $.
Agents' first order conditions are necessary and sufficient and they imply for each agent $h$ and all $ t=0,\ldots,T-1 $ that
\begin{equation} c_{h,t+1}=c_{h,t}\,(\beta ( 1+r_{t+1}))^{1/\gamma_h} . \end{equation}
By market clearing $ K_{t+1}=Y_t-\sum_{h} c^h_t$ for all $ t$.
An equilibrium then consists of $(c^h_t)^{h \in {\mathbf H}}$

\begin{eqnarray}
\label{shoot1}
&&  c_{h,t+1}=c_{h,t}\,(\beta ( 1+r_{t+1}))^{1/\gamma_h}, \ t=0,\ldots T-1, h \in {\mathbf H} \\
\label{shoot2}
&& K_{t+1}=Y_t-\sum_{h \in {\mathbf H}} c^h_t, t=0,\ldots, T, \  K_{T+1} = 0\\
\label{shoot3}
&& p_t=\beta^t \frac{u'_1(c_t)}{\sum_{s=0}^T \beta^s  u'_1(c_s)} \ \  \forall t\\
\label{shoot4}
&& \sum_t p_t (c^h_t -w_t \ell^h_t-d_h \tau_t e_t) = p_0 (1+r_0) k^h_0 
\end{eqnarray}
Clearly initial $ (c^h_0)_{h \in {\mathbf H}} $ uniquely determine a terminal $ K_{T+1} $ and all paths $ (c^h_t) $. As we see below, it will be useful to write future capital, prices, emissions and consumption as a function of initial aggregate consumption, $ C_0 $ and Negishi weights $ (\lambda_h)_{h \in {\mathbf H}} \in \Delta^{H-1}$, 
$ K_t(\lambda,C_0)$, $ p_t(\lambda,C_0) $ etc.

\subsection{An example with multiple equilibria}
                                                                               
We consider 
a $ 5 -$period  \lq\lq business-as-usual\rq\rq \ (BAU, $ \tau_t=0 \ \forall t$) climate--economy model with
two four-period-lived agents who differ only in their coefficient of relative
risk aversion, $\gamma_1 = 1.5$ and $\gamma_2 = 2.5$.  There is no carbon tax
and no abatement; the climate--damage feedback is active.

With cumulative emissions $\bar e_t$ and temperature $T_t=1.7\,\bar e_t$,
\begin{align*}
  \text{damage:}\quad & D_h(T)=\min\{0.02 \cdot T^{3},0.99\}, h=1,2\\
  \text{depreciation:}\quad & \delta(T)=\min\{0.1+0.02 \cdot T^{3},\ 0.99 \}.
\end{align*}
Emissions are a single spike,
$\kappa_2=1.0$ and $\kappa_t=0$ otherwise.
We take $ \alpha=0.33 $
, $\beta=0.97$, $ \bar \ell_1=\bar \ell_2=0.5$ and $ \bar e _0=0.5 $.

For these parameters, the terminal-capital condition $k_T(\lambda_1,c_0)=0$ has
\emph{three} branches in $c_0$ . Table \ref{tab:equilibria} shows the allocations for the three equilibria.  On each branch there is a unique
Pareto weight $\lambda_1$ that clears both budgets.  This yields three distinct
competitive equilibria, each satisfying $k_T=0$ and
$\mathrm{budget}_1=\mathrm{budget}_2=0$ to machine precision (penalty
$\sim 10^{-24}$). Note that similar examples can be constructed with a single agent - multiplicity of equilbria in this economy thus arises because of the externality (see \cite{foster1970price}  or \cite{kehoe1992characterizing} for early examples with the same economic interpretation).

\begin{table}[h]

\centering
\small
\begin{tabular}{lllllll}
\toprule
 & & $t=0$ & $t=1$ & $t=2$ & $t=3$ & $t=4$ \\
\midrule
eq1 & $k_t$    & $3.974$ & $4.708$ & $5.455$ & $6.209$ & $0.248$ \\
    & $c_{1,t}$& $0.299$ & $0.294$ & $0.287$ & $0.038$ & $0.005$ \\
    & $c_{2,t}$& $0.085$ & $0.084$ & $0.082$ & $0.025$ & $0.007$ \\[2pt]
eq2 & $k_t$    & $3.974$ & $4.286$ & $4.612$ & $4.953$ & $0.200$ \\
    & $c_{1,t}$& $0.589$ & $0.582$ & $0.572$ & $0.083$ & $0.018$ \\
    & $c_{2,t}$& $0.216$ & $0.214$ & $0.212$ & $0.067$ & $0.027$ \\[2pt]
eq3 & $k_t$    & $3.974$ & $3.695$ & $3.416$ & $3.129$ & $0.358$ \\
    & $c_{1,t}$& $0.848$ & $0.844$ & $0.844$ & $0.253$ & $0.122$ \\
    & $c_{2,t}$& $0.549$ & $0.547$ & $0.547$ & $0.265$ & $0.172$ \\
\bottomrule
\end{tabular}
\caption{Equilibrium trajectories.  In every equilibrium capital is wiped out
between $t=3$ and $t=4$ by the $t=2$ emission spike acting through the cubic
depreciation feedback; the equilibria differ in how much capital is
accumulated beforehand and hence in consumption levels and long-run
temperature. In all equilibria $ k_5=0$ up to pachine precision.}
\label{tab:equilibria}
\end{table}

The three equilibria are \emph{Pareto-ranked} and, here, also
\emph{temperature-ranked}: the light-saving equilibrium eq3 has the highest
welfare \emph{and} the lowest long-run temperature, while the heavy-saving
equilibrium eq1 is both the poorest and the hottest.
Heavier saving (eq1) builds more capital, emits more at the
spike, ends hotter and poorer.  Note that in eq1/eq2 the less risk-averse
agent~1 consumes far more than agent~2, reflecting the large Pareto weight
$\lambda_1$ that clears the budgets

\subsubsection{Bayesian optimization}
We use BO with UCB acquisition function and SE kernel so solve (\ref{climprob}) for this simple calibration.
We take as the planner's objective simply the sum of utilities and assume that carbon taxes must be zero. The planner simply chooses the best competitive equilibrium.

\begin{figure}[h]
\centering
  \includegraphics[width=\linewidth]{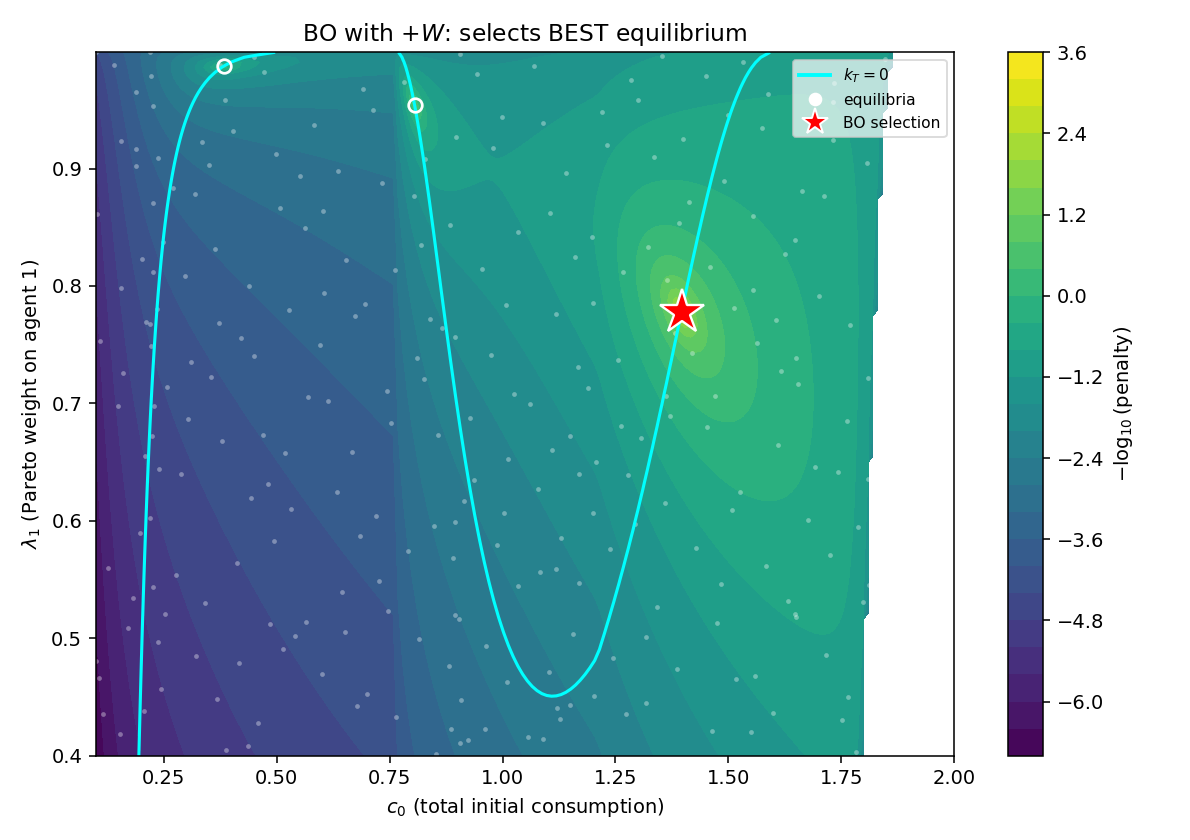}
\caption{Bayesian-optimization selection of equilibria for
$\gamma=(1.5,2.5)$.  Cyan: $k_T=0$;  three competitive equilibria.  Adding $+W$  to the
objective makes the previously three-fold-degenerate optimum unique, and BO
converges to the best equilibrium.}
\label{fig:bo}
\end{figure}

As one would expect from the example in the previous section, this is a relatively difficult problem for BO. Nevertheless, the method correctly identifies the best of the three competitive equilibria as the solution. Figure \ref{fig:bo} shows a contour plot of the objective function. The $ K_T=0 $ line is plotted for convenience (not an output of BO). In the figure, three local optimia can be identified. The global optimum comes with high initial consumption and (relatively) low emissions.
The star marks the equilibrium selected by BO.

As above, the example is mainly illustrative, other numerical methods can be used to identify the three equilibria. However, the example shows that BO is a competitive method. More importantly, the example shows that there is a possibility of multiple equilibria in simple climate macro models. While the calibration in this example is rather extreme, it is an important quantitative question if multiplicity is possible in models with more standard calibrations for damages and emissions. BO is ideally suited to address this question and in larger scale problems might be the only numerical method available for this task.

The example demonstrates that in a model with externalities, multiplicity of equilibria can occur naturally. However, the example is obviously extreme  in that the assumed damage functions and emissions path are very far from what is considered realistic in the literature. 

\subsection{A feasible strategy to use BO}

In order to apply BO efficiently to realistic large-scael calibration, we want to formulate the problem as an unconstrained problem and penalize the constraint-violations. For finite $T$, a naive approach would be to use BO to solve the following optimization problem.

\begin{eqnarray}
    \label{climprob}
&& \max_{\substack{(\tau_t) \in {\mathbb R}^{T+1}_+, d \in \Delta^{H-1},\\ C_0 >0, \lambda \in \Delta^{H-1}}}    W\left( (c^h_t(\lambda,C_0,\tau)^{h \in {\mathbf H}}_{t=0\ldots T}\right) - \eta_1 K_{T+1}(\lambda,C_0,\tau) ^2- \eta_2 \sum_{h \in {\mathbf H}} \\ \notag
&&  \left( \sum_t p_t(\lambda,C_0,\tau) (c^h_t(\lambda,C_0\tau) -w_t(\lambda,C_0,\tau) \ell^h_t(\lambda,C_0,\tau) - d_h \tau_t e_t(\lambda,C_0,\tau) ) -  p_0(\lambda,C_0,\tau)(1+r_0) k^h_0 \right)^2
\end{eqnarray}

Unfortunately, while budget violations have clean economic interpretations, violations of the terminal condition in capital, $ K_{T+1}=0 $ do not. In practice, one could try to choose $ \eta_1 $ sufficiently large so that, at the optimal solution, $ K_{T+1} $ gets pushed very close to zero. Because the object certified by the BO algorithm is the objective of finite penalties rather than the original constrained Ramsey problem, the scale of $\eta_1$ and $\eta_2$, the sensitivity to these choices, and the specific penalty contributions at each optimum are integral to interpreting the results. As $\eta$ grows, the objective can become steeper, directly inflating the Lipschitz constants on which the certification argument depends. 
Moreover, it is well known that for large $T$, forward shooting is not a reliable numerical method. Small errors in the initial $ (c_0^h) $ propagate quickly over time, and capital can quickly become negative. In that sense, the formulation (\ref{climprob}) can only be useful for small $T$. 

To tackles large-scale problems efficiently, we therefore choose a different approach and proceed in two steps. First, we establish that for the calibration we consider, with high probability, the initial $ C_0 $ is uniquely determined by the firm's first order conditions, for any given $ \lambda \in \Delta^{H-1}$ and given policy. 
Then we use Newton's method to solve for the capital path, given $ \lambda $ and obtain a BO problem only in policy-variables (taxes and transfers) and Negishi weights. This second step assumes the uniqueness of the capital path, i.e. it is only sensible with high probability.

\subsubsection{Probably monotone shooting}
For given $ (\tau_t)_{t=0}^{T} $ and $ \lambda \in \Delta^{H-1}$, initial consumption $ C_0 $ uniquely determines the path of capital. An equilibrium is characterized by $ K_{T+1}=0 $.
It is useful to reformulate (\ref{shoot1})-(\ref{shoot4}) and follows.
\begin{align}
k_{t+1} &= \Phi\!\bigl(y(k_t,\bar{e}_t,\tau_t,t) - C_t\bigr),\label{eq:kfwd}\\
c_{h,t+1} &= \underline{c}\bigl(c_{h,t}\,(\beta\,(1+r_{t+1}))^{1/\gamma_h}\bigr),
\qquad C_{t+1}=\textstyle\sum_h c_{h,t+1},\label{eq:cfwd}\\
\bar e_{t+1} &= \bar e_t + e(k_t,\bar{e}_t,\tau_t,t).\label{eq:sfwd}
\end{align}
Here $\Phi$ (.) is a smooth lower bound on capital and
$\underline{c}$ (.) a smooth lower bound on consumption.\footnote{In our implementation below, both belong
to the one--parameter family of \emph{softplus floors}: writing
$\varsigma(z)=\log(1+e^{z})$ for the softplus and $\varrho(z)=\varsigma'(z)=1/(1+e^{-z})$
for the logistic sigmoid, define
\begin{equation}
\mathcal F(v;\,a,s) \;=\; a + s\,\varsigma\!\Bigl(\frac{v-a}{s}\Bigr)
\;=\; a + s\,\log\!\Bigl(1+e^{(v-a)/s}\Bigr),
\qquad
\mathcal F'(v;\,a,s) = \varrho\!\Bigl(\frac{v-a}{s}\Bigr),
\label{eq:softfloor}
\end{equation}
which is $C^\infty$, satisfies $\mathcal F(v)\to v$ as $v\to+\infty$ and
$\mathcal F(v)\to a$ as $v\to-\infty$, so it acts as a kink--free
$\max\{v,a\}$ with smoothing scale $s$. The two instances are
$
\Phi(v) = \mathcal F\!\bigl(v;\ \tfrac{\underline k}{2},\ \underline k\bigr)
    $ and $ \underline c(v) = \mathcal F\!\bigl(v;\ \underline c_{\min},\ \underline c_{\min}\bigr)$ .}
 These
floors keep $k_t,c_t$ strictly positive without introducing kinks, which is
essential both for differentiability.
The new system (\ref{eq:kfwd})-(\ref{eq:sfwd}) allows to consider the function $ K_{T+1}:(0,Y_0) \rightarrow {\mathbb R} $ and examine its zeros. If this function is monotone, it must have a unique zero, and for given $ \tau $ and $ \lambda $ 
the equilibrium is unique. In this case, we say shooting is monotone.
It is easy to see that the derivative $ \frac{ d K_{T+1}}{dC_0} $ can be computed analytically. Hence, we can use the methods from \ref{probmonfun} to establish uniqueness for all $ \lambda \in \Delta^{H-1} $ and all taxes. In fact, for this one dimensional case the method simplifies since we can use directly the derivative and do not have to evaluate a function of the eigenvalues of a Jacobian.
\subsubsection{Optimal policy under unique capital paths}
After establishing that forward shooting is monotone with high probability, we can replace it with Newton's method, that is to say, we solve (\ref{eq:kfwd})-(\ref{eq:sfwd}) as one large non-linear system of equations using standard non-linear solvers.
All variables are (smooth) function of $ \lambda $ and $ \tau$ and we can write the planner's problem as follows.
\begin{eqnarray}
    \label{climprob}
&& \max_{\substack{(\tau_t) \in {\mathbb R}^{T+1}_+, d \in \Delta^{H-1},\\ \lambda \in \Delta^{H-1}}}    W\left( (c^h_t(\lambda,\tau))^{h \in {\mathbf H}}_{t=0\ldots T}\right) - \eta \sum_{h \in {\mathbf H}}\\ \notag
&&   \left( \sum_t p_t(\lambda,\tau) (c^h_t(\lambda,\tau) -w_t(\lambda,\tau) \ell^h_t(\lambda,\tau) - d_h \tau_t e_t(\lambda,\tau))  -  p_0(\lambda,\tau)(1+r_0) k^h_0 \right)^2.
\end{eqnarray}

We can then use BO to find a solution and certify the solution probabilitically. To illustrate this we consider one specific example.

\subsection{A simple large scale calibration}
We take $ T=\infty $, $ H=4 $ and consider the following specification of damages. 
\begin{align*}
  \text{high damage:}\quad & D_h(T)=\min\{ 1.6 \cdot 0.015 \cdot T^{2},0.99\}, h=1,2\\
   \text{low damage:}\quad & D_h(T)=\min\{ 0.4 \cdot 0.015 \cdot T^{2},0.99\}, h=3,4\\
  \text{depreciation:}\quad & \delta(T)=\min\{0.1+0.02 \cdot T^{2},\ 0.99 \}.
\end{align*}
The emission intensities are 
$\kappa_t=0.1$ for $ t \le 5 $ and $\kappa_t=0.1 (10-t)/10 $ for $ 5< t\le 10 $, $ \kappa_t=0$ otherwise.
We take $ \alpha=0.33 $
, $\beta=0.97$, $ \bar \ell_h=\frac{1}{4} $, $ k^h_0=\frac{1}{4} $  for all $h$, and $ \bar e _0=0.9 $.
We assume $ \gamma_1=\gamma_3=2$ and $ \gamma_2=\gamma_4=3$.
We assume that the social planner has a strong preference for consumption equality and that $ \gamma^P=5 $.
As the abatement technology-parameters we choose $ \theta_1=0.5 $ and $\theta_2=2.5$.

Overall, the calibration  is in line with values used in the literature (e.g. \cite{kubler2025using}). The two aspects that are somehwat unusual is that emissions are extremely high for 5 periods and then quickly decline to zero and that damages are much higher than in the Norhaus specifiction (\cite{barrage2024policies}). The fact that emissions are zero after 10 periods is motivated by the need for computational feasibility. The specification of damages is somewhat at the extreme end of specifications found in the literature  (see, e.g., the \lq\lq 3X calibration\rq\rq \ in \cite{kotlikoff2021making}).

\subsubsection{Unique capital paths}
\label{sec:promoncap}
As explained above, the first step of our numerical methods establishes that, with high probability, capital paths are unique for all Negishi-weights and admissible sequences of taxes. In our main exercise below, we assume a constant tax rate over time (and then compare the results with more flexible schemes). For tractabiltiy, we focus on this case in this exercise, i.e. we assume that $ \tau_t = \bar \tau \ge 0$ for all $t$ and establish probable uniqueness for all $ \bar \tau $ and all $ \lambda $.

The method so far has focused on the case where $T$ is relatively small, and here we want to apply it to an infinite horizon model. An important property of our calibration is that while $ T $ is infinite, $ \kappa_t = 0 $ for all $ t \ge T_S $, with $ T_S $ being relatively small.  It is easy to see that given $ K_{T_S} $ and $ \bar e_{T_S} $, there is a unique continuation for the capital path starting at $ T_S.$
We therefore split the time horizon into two segments.
The shoot segment $ t=0,\ldots,T_{\mathrm{split}}-T_s $ computes a capital path for given initial $ C_0$ and outputs $k^{\mathrm{fs}}_{T_{\mathrm{split}}+1}$.
The BVP segment ($t=T_{\mathrm{split}},\dots,\infty$) solves the households' Euler system exactly by root--finding on
the interior capital unknowns $(k_{T_{\mathrm{split}}+1},\dots,k_{T_\infty})$,
with boundary conditions $k_{T_{\mathrm{split}}}=k_{\mathrm{split}}$  (handed over
from the shoot) and $k_{T_{\infty}}=k_{T_{\infty}-1} $  for some sufficiently large $ T_{\infty} $. We denote new capital at $ T_{\mathrm{split}} $ computed by Newton's method by
$k^{\mathrm{bvp}}_{T_{\mathrm{split}}+1}$ and the capital choice implied by the shoot by  $ k^{\mathrm{fs}}_{T_{\mathrm{split}}+1}$ and consider the function
\begin{equation}
\Delta_k(C_0) \;=\; k^{\mathrm{fs}}_{T_{\mathrm{split}}+1} \;-\; k^{\mathrm{bvp}}_{T_{\mathrm{split}}+1},
\label{eq:deltak}
\end{equation}
which gives the discrepancy at the first post--split period between the forward--shot capital
and the BVP capital. In equilibrium, $\Delta_k=0$.
We show that for all $ \lambda, \bar \tau $, $ \Delta_k $ is monotone with high probability.

With $256$ Sobol seeds and $100$ UCB iterations, BO identifies  $f=-1.227345$ as the largest derivative on the domain. The 
covering bounds give that conditional on the Lipschitz bound being correct,
\[
\mbox{Prob}\Bigl(\sup_{x\in\mathrm{box}} f(x)\ge 0\Bigr)\ \le\ 10^{-8}
\]
The exceedance probabilities, computed from the PI acquisition function are so tiny that $\Phi$
underflows, so the bound is dominated by the Lipschitz confidence. Taking $\delta_L=0.01$ results in large Lipschitz bounds, but the resulting covering numbers are still well below $ 10^{10} $, resulting in an overall probabilities around 0.01.

\subsubsection{BAU scenario}
Although capital paths are unique for all $ \lambda $, multiple equilibria are still possible -- after all $ \lambda $ is determined by the budget balance requirement for all agents. 
We want to investigate the potentiality of multiple equilibria in the business as usual (BAU) case without policy.
In this case, the externality is strongest. We take the planners welfare function to be zero and simply maximize  the sum of the (negative) penalty terms\footnote{As in Section \ref{sec:probmonex} one could also try to establish monotonicity for this case, the chosen approach seems more straightforward and better suited for the problem at hand.}.

 The BO algorithm finds an optimum at 
$ \lambda=(0.34, 0.06, 0.49, 0.11) $ with a value of the objective function of $ -0.0009 $. 
We establish Lipschitz bounds that hold with probability at least 0.99. These bounds lead to the astronomical covering number of $ 2 \times 10^{12}$. 
Nevertheless, it can be established that under the established Lipschitz bound the probability that outside of a (sup-norm) radius of 0.05 for $ \lambda $, the value of the objective is below $ -0.002 $ with probability $ 1-10^{-11} $. Unlike in the simple 5 period example above, all functions seem extremely well behaved and the objective is single peaked, decreasing rapidly away from the identified optimum. With high probability, there is a unique BAU equilibrium for this calibration.
In this context, it is not a meaningful statement to establish that with high probability there is no point that gives a function value significantly (i.e. more than 0.002) higher than the best-observed value since by construction the function cannot exceed zero.

At the new long-run steady-state the temperature is 3.12 degrees above pre-industrial average, resulting in large damages and an aggregate consumption at the steady state of around 0.73.

\subsubsection{Optimal carbon taxes}
In general, optimal carbon taxes are a sequence of taxes $ (\tau_t) $ over all periods with $ \kappa_t > 0 $.
In our calibration, it is feasible to compute the entire optimal sequence (since $\kappa_t=0$ for $ t\ge 10 $).
However, it is not feasible to certify the solution with high probability. The covering bounds from Section \ref{sec:regret-bounds} obviously face a severe curse of dimensionality.
With ten carbon taxes (and three Negishi weights and initial aggregate consumption) this becomes a fairly large BO problem. Going much beyond 20 periods of positive carbon taxes, the problem quickly  becomes infeasible. The alternative is to assume that taxes change linearly (or quadratically) over time. This reduces the computational burden to 2-3 policy variables instead of 20. In the context of optimal carbon taxes, there is evidence that  (see, e.g. \cite{hassler2018consequences}) small variations in the tax have little effect on welfare. This is confirmed in our calibration, where we consider taxes that are constant over time. In general, however, there is no guarantee that simple function forms provide a good approximation to the optimal sequence.

We consider a constant tax and a fully flexible tax.
In both examples, the carbon tax revenue is large enough to ensure that the budget penalties are not binding. The optimization problem simplifies quite clearly, since the planner simply chooses taxes as instruments, the resulting Negishi-weights have the property that each agents' budget balance can be achieved with the right level of carbon dividends $ d \in \Delta^{H-1} $. The penalty term becomes irrelevant.
\begin{table}[h]
\centering
\small
\begin{tabular}{lllll}
  &  BAU  & Constant  & Time dependent   \\ \hline
  ind. welfares  &  [0.18, 0.18, 0.22, 0.22] &   [0.24, 0.24, 0.24, 0.24]  & [0.24, 0.24,  0.24, 0.24] \\ 
 Ss agg. cons. & 0.73   & 0.95   & 0.95\\
 Ss temp. &   3.12    & 1.99 &  1.98 \\
  SWF & 0.139  & 0.168 &  0.169 \\
  d  &  NA &       [0.41, 0.40, 0.10, 0.09] &   [0.39, 0.41, 0.1, 0.1]  \\ 
\hline
\end{tabular}
\caption{Key properties of optimal solutions for different tax schemes}
\label{tab:optimalt}
\end{table}
Table \ref{tab:optimalt} shows that individual welfares and long-run aggregate consumption increase greatly with carbon taxation (comparing BAU to any other column) but barely change between the two different tax schemes. Even a constant carbon tax reduces long run temperature dramatically (from 3.21 degrees to 1.99 degrees) and reduces damages (that are quadratic in temperature) dramatically. The values of the social welfare function  are much larger in the cases with tax-taxes  than without, but again the actual tax scheme makes little difference.

Note that the presence of carbon tax revenue allows the planner to redistribute between agents that are less affected by climate change and those that are affected more. The welfare of agents 1 and 2 increases greatly, not only because emissions are lowered but also because of these implicit transfers.
For different social welfare functions, the transfers can be much lower and the improvement via carbon taxation becomes smaller.

As in the BAU example, also in the examples with constant taxes it can be established that with high probability the BO algorithm finds the correct solution. In this context, it is a meaningful statement to establish that with a probability of almost 99 percent (the Lipschitz bound being correct with the probability of 99 percent and the probability that the GP does not exceed the specified value being $ 10^{-6}$) the value of the objective function is not more than 0.5 percent above the identified optimal value.

\section{Conclusion}

Many economic applications of interest give rise to intractable non-convex optimization problems. This paper argues that they can be tackled with Bayesian optimization. Although the optimum cannot be located with certainty, it can be approximated with high probability. The Bayesian optimization literature provides a decision-theoretic foundation for this approximation by reducing the problem to a bandit problem. Establishing how that foundation can be micro-founded within the optimal policy problems arising in economics is left to future work.

\bibliographystyle{ecta}
\bibliography{bib_econ_clean}
\renewcommand{\appendixpagename}{Appendix}
\begin{appendices}

\section{Additional results on Gaussian Processes} \label{app:sec2}

\subsection{Alternative kernels}
Although the examples in this paper all use the SE-kernel it is useful 
to also consider an alternative class of kernels. Whenever the objective function is not $ C^{\infty}$, the use of the SE-kernel is not appropriate.
The Mat\'ern kernel that is widely used in Gaussian process regression, was introduced by Bertil Mat\'ern in 1960 (see \cite{matern2013spatial}). It generalizes the squared exponential kernel by introducing a smoothness parameter $\nu > 0$ that controls the differentiability of the resulting sample paths. The general form is given by
\[
  k_\nu(r) = \frac{2^{1-\nu}}{\Gamma(\nu)}
  \left(\frac{\sqrt{2\nu}\, r}{\ell}\right)^{\!\nu}
  K_\nu\!\left(\frac{\sqrt{2\nu}\, r}{\ell}\right),
\]
where $r = \|\mathbf{x} - \mathbf{x}'\|$ is the Euclidean distance between inputs, $\ell > 0$ is the length-scale hyperparameter, $\Gamma(\cdot)$ is the gamma function, and $K_\nu$ is the modified Bessel function of the second kind of order $\nu$. A Gaussian process with Mat\'ern-$\nu$ covariance has sample paths that are $\lceil \nu \rceil - 1$ times mean-square differentiable.
 
When $\nu$ takes the half-integer values $\nu = p + \tfrac{1}{2}$ for $p \in \mathbb{N}_0$, the kernel simplifies to a product of an exponential and a polynomial, making it computationally convenient. The two most commonly used special cases are as follows.
\begin{itemize}
    
\item Mat\'ern $\tfrac{3}{2}$:
 Setting $\nu = \tfrac{3}{2}$ yields sample paths that are once mean-square differentiable and we can write the kernel as
\[
  k_{3/2}(r) = \left(1 + \frac{\sqrt{3}\,r}{\ell}\right)
  \exp\!\left(-\frac{\sqrt{3}\,r}{\ell}\right).
\] 
\item Mat\'ern $\tfrac{5}{2}$: Setting $\nu = \tfrac{5}{2}$ yields sample paths that are twice mean-square differentiable and we can write the kernel as
\[
  k_{5/2}(r) = \left(1 + \frac{\sqrt{5}\,r}{\ell}
  + \frac{5\,r^2}{3\ell^2}\right)
  \exp\!\left(-\frac{\sqrt{5}\,r}{\ell}\right).
\]
The Mat\'ern $\tfrac{5}{2}$ kernel is particularly popular in Bayesian optimization, where its moderate smoothness assumption often matches the regularity of real-world objective functions more faithfully than either Mat\'ern $\tfrac{3}{2}$ or the squared exponential.
\end{itemize} 
As $\nu \to \infty$, the Mat\'ern kernel converges to the squared exponential (RBF) kernel $k_{\mathrm{SE}}(r) = \exp\bigl(-r^2/(2\ell^2)\bigr)$, which corresponds to infinitely differentiable sample paths. At the other extreme, $\nu = \tfrac{1}{2}$ recovers the Ornstein--Uhlenbeck kernel $k_{1/2}(r) = \exp(-r/\ell)$, whose sample paths are continuous but nowhere differentiable. 

In our examples, one can also use a Mat\'ern 5/2 kernel and results turn out to be very similar, with slightly more sample points needed to establish the probabilistic bounds.
\subsection{Reproducing Kernel Hilbert Spaces}
\label{rkhs}

The main assumption underlying the probabilistic bounds is that the true objective function is a sample path of the Gaussian process. This is typically not exactly true but holds as an arbitrarily good approximation.
In order to investigate the relationship between the objective function and Gaussian process regression formally, it is necessary to define the concept of a reproducing kernel Hilbert space (see, e.g. \cite{kanagawa2025gaussian} for a survey).
\begin{definition}[Reproducing Kernel Hilbert Space]
Let $\mathcal{H}$ be a Hilbert space of real-valued functions on a set $\mathcal{X}$ with inner product $\langle \cdot, \cdot \rangle_{\mathcal{H}}$. We say $\mathcal{H}$ is a \textbf{reproducing kernel Hilbert space} (RKHS) if there exists a function $k: \mathcal{X} \times \mathcal{X} \to \mathbb{R}$ such that:
\begin{enumerate}
\item For all $x \in \mathcal{X}$, $k(\cdot, x) \in \mathcal{H}$
\item  For all $f \in \mathcal{H}$ and $x \in \mathcal{X}$,
\begin{equation}
f(x) = \langle f, k(\cdot, x) \rangle_{\mathcal{H}}
\end{equation}
\end{enumerate}
The function $k$ is called the \textbf{reproducing kernel} of $\mathcal{H}$.
\end{definition}

The Moore-Aronszajn theorem states that
for every positive definite kernel $k$ on $\mathcal{X} \times \mathcal{X}$, there exists a unique RKHS $\mathcal{H}_k$ of functions on $\mathcal{X}$ for which $k$ is the reproducing kernel. Conversely, every RKHS has a unique reproducing kernel (see, e.g. \cite{williams2006gaussian}).
This implies that by selecting a covariance kernel of the GP one directly selects an associated RKHS.
We can explicitly construct the RKHS as the closure of the following 
pre-Hilbert space. Let
\[
\mathcal{H}_0 = \left\{ \sum_{i=1}^n c_i\, k(\cdot, x_i) \;:\; 
n \in \mathbb{N},\; c_1,\ldots,c_n \in \mathbb{R},\; 
x_1,\ldots,x_n \in \mathcal{X} \right\}
\]
with an inner product between any two elements $f, g \in \mathcal{H}_0$,
\[
f := \sum_{i=1}^n a_i\, k(\cdot, x_i), \qquad 
g := \sum_{j=1}^m b_j\, k(\cdot, y_j) \quad \text{defined as}
\]
\[
\langle f, g \rangle_{\mathcal{H}_0} 
= \sum_{i=1}^n \sum_{j=1}^m a_i b_j\, k(x_i, y_j).
\]
It is useful to note that the resulting RKHS norm $ \|\cdot \|_{\mathcal H_k}$ of a function $f$ captures is magnitude (like e.g. the $ L_2 $ norm) but also its smoothness. The smaller the norm, the smoother the function.

In GP regression, the posterior mean lies in the RKHS since it can can be written as:
\begin{equation}
\mu(x) = \sum_{i=1}^n \alpha_i k(x, x_i)
\end{equation}
where $\alpha = (K + \sigma_{\epsilon}^2 I)^{-1} y$ for observations $y$ at points $x_1, \ldots, x_n$. More interestingly, for the case of no noise, i.e. $ \sigma^2_{\epsilon}=0$, the difference between the posterior mean and the true function $f$ satisfies the following bound that uses the posterior variance (see, e.g. Theorem 4.8 in \cite{hennig2022probabilistic}).
\begin{equation}
\sup _{f \in {\mathcal H}_k, \| f \|_{{\mathcal H}_k} \le 1} (\mu(x)-f(x))^2 = \sigma(x)^2
\mbox{ for all } x \in {\mathcal X}    
\end{equation}
Note that for an asymptotic analysis, this bound turns out to be very important, but that in our finite sample setting it is of rather limited use since we do not know the kernel norm of the true function -- it does make clear however, that any function in the RKHS of a Gaussian process can be efficiently approximated by the posterior mean.

In our setting, the crucial question is 
when can we expect functions that arise in economic problems to be a sample path of the specified Gaussian process. This is the main assumption that underlies much of the analysis in BO  but, unfortunately, does not have a clear-cut      and simple  answer. We first consider necessary conditions, then explain why sufficient conditions are difficult to verify, and finally show that the statement is true in a satisfactory approximate sense if one considers the SE kernel.

Although it turns out that given an RKHS, one can construct continuous functions on compact domains that do not lie in the RKHS, \cite{scholpple2026spaces} prove the following result (in their Theorem 22).
\begin{lemma}
\label{lemma1}
Let $\Omega \subset \mathbb{R}^d$ be a bounded domain with smooth boundary and let
$C^k(\Omega)$ denote the space of $k$-times continuously differentiable functions on
$\Omega$.  Then there exists a reproducing kernel
Hilbert space (RKHS) $H$ with a bounded kernel such that
$$
  C^k(\Omega) \;\subset\; H \;\subset\; C_0(\Omega)\,
$$
 if and only if
\[
  k \;>\; \frac{d}{2}.
\]
\end{lemma} 
Moreover, for the Mat\'{e}rn kernel the following result can be derived from Lemma \ref{lemma1} and \cite{steinwart2024does}.
\begin{lemma}
Let $T \subset \mathbb{R}^d$ be a bounded domain, $k \in \mathbb{N}$, and $X$ be
a centered Gaussian process with Mat\'ern-$\nu$ covariance kernel. There exists
an RKHS $H$ with bounded kernel such that
\[
  C^k(T) \;\subset\; H \qquad \text{and} \qquad P(Y_\bullet \in H) = 1
\]
if and only if
\[
  \nu > k \qquad \text{and} \qquad k > \frac{d}{2}.
\]
\end{lemma}
Clearly, the lemma cautions us to use a kernel that is not sufficiently smooth in high-dimensional problems. As the dimension of our BO problems remains below 5, a Mat\'ern 5/2 kernel is appropriate, beyond that a smoother kernel must be used. However, it does not justify the assumption that the true function is a sample path of a GP.

To discuss sufficient conditions \cite{karvonen2023small} introduces the "sample support set", the largest set contained in every element of full measure under the law of the GP in the $\sigma$-algebra induced by the collection of scaled RKHSs. This potentially non-measurable set consists of those functions that can be expanded in terms of an orthonormal basis of the RKHS of the covariance kernel of the GP and have their squared basis coefficients bounded away from zero and infinity. (\cite{karvonen2023small}. For the SE kernel, the natural orthonormal basis of its RKHS  consists of Hermite functions: using the Gaussian measure, the eigenfunctions of the Gaussian kernel are Hermite functions. The Karhunen–Lo\'eve expansion of the GP with the SE kernel expands in this basis, with eigenvalues decaying super-exponentially. A  $C^{\infty}$ function $f$ is in the sample support set if and only if its Hermite-function expansion coefficients $(c_n) $ satisfy that there exist constants $ 0 < c \le  C < \infty $  such that $c \le  |c_n|²/\lambda_n \le C $ for all $n$, where $ (\lambda_n) $ are the (super-exponentially decaying) eigenvalues of the SE kernel's integral operator. This is an extremely strong constraint — it requires the coefficients to decay at precisely the right super-exponential rate.

While it is difficult to show that the true function is a sample path, it is straightforward to show that there are functions arbitrarily close (in the sup-norm) to the true function that are a sample path of our GP with SE kernel.

\cite{van2008reproducing} prove that
the support of a mean-zero Gaussian random element  in a separable Banach space $B$ is the closure of its RKHS in $B$.
For a GP with the SE kernel defined on a compact domain, taking $B$  to be the space of continuous functions with the sup-norm, the RKHS of the SE kernel consists of analytic functions with a specific decay rate on their Fourier (or Hermite) coefficients. So the topological support is the closure of this RKHS in the sup-norm topology.
The practical consequence (see, e.g. \cite{van2011information}) is that the probability of a draw of the Gaussian process falling in a neighbourhood of a given continuous $ f $ is typically positive, no matter how small the neighbourhood. This means every  continuous function in the interior of the sup-norm closure of the RKHS is in the support. For the SE kernel on a compact domain, the RKHS is dense in the space of continuous function so the support in the sup-norm topology is all of C(T) — meaning every continuous function has positive probability of being approached arbitrarily closely. 

Obviously, this does not imply that every $ C^\infty$ function is a sample path in the almost-sure sense -- however, for our purposes this approximation is sufficient. First, in all numerical work functions are computed with machine precision, and the distinction between an arbitrary good approximation to the true function and the true function itself becomes meaningless. Second, since we are only interested in the maximum of the true function, it is clear that this also can be arbitrarily well approximated by the maximum of a sample path of the GP.  Note that in order for our Lipschitz certificates are to apply beyond the exact Bayesian model, the approximation error does need to enter the probability or regret statements, since there are sample paths for which this error is arbitrarly small.

\subsection{Hyperparameters}
Given training data $\{(\bx_i, y_i)\}_{i=1}^{n}$ with inputs $\bx_i \in \R^d$
and noisy scalar targets $y_i$, we model the latent function $f$ as a
Gaussian process,
\[
f \sim \mathcal{GP}\bigl(0,\, k(\bx, \bx')\bigr),
\qquad
y_i = f(\bx_i) + \varepsilon_i,
\quad
\varepsilon_i \sim \mathcal{N}(0, \sigma_n^2).
\]
The squared-exponential (SE, or RBF) kernel is
\[
k(\bx, \bx') \;=\; \sigma_f^2
\exp\!\left(-\frac{\lVert \bx - \bx' \rVert^2}{2\,\ell^2}\right),
\]
with \emph{hyperparameters} collected in
$\bm{\theta} = (\sigma_f^2,\, \ell,)$:
 signal variance $\sigma_f^2$, length-scale $\ell$ (a single
value for the isotropic case; one per dimension $\ell_1,\dots,\ell_d$ for
automatic relevance determination). Estimation of $\bm{\theta}$ from the data is of  central importance for the performance of BO.

\subsubsection{The marginal likelihood}

Let $K \in {\mathbb R}^{n\times n}$ have entries $K_{ij} = k(\bx_i, \bx_j)$ and define
$\Ky = K + \sigma_n^2 I$. Marginalising over the latent values $f$ gives a
Gaussian distribution over the observed targets, and its log density---the
\emph{log marginal likelihood}---is
\[
\log p(\by \mid X, \bm{\theta})
= -\tfrac{1}{2}\, \by^\top \Ky^{-1} \by
  \;-\; \tfrac{1}{2}\log\lvert \Ky \rvert
  \;-\; \tfrac{n}{2}\log 2\pi .
\]
The three terms have intuitive roles: the first rewards fitting the data,
the second penalises model complexity, and the last is a normalising
constant. Maximising this quantity with respect to $\bm{\theta}$ is the
standard estimation procedure, often called \emph{type-II maximum
likelihood} or \emph{empirical Bayes}, because the complexity penalty
provides an automatic Occam's-razor balance.

Gradient-based optimisation is efficient because the derivatives are
available in closed form. Writing $\bm{\alpha} = \Ky^{-1}\by$, the
derivative with respect to any hyperparameter $\theta_j$ is
\[
\frac{\partial}{\partial \theta_j}
\log p(\by \mid X, \bm{\theta})
= \tfrac{1}{2}\,
\mathrm{tr}\!\left(
\bigl(\bm{\alpha}\bm{\alpha}^\top - \Ky^{-1}\bigr)
\frac{\partial \Ky}{\partial \theta_j}
\right).
\]
The kernel derivatives $\partial \Ky / \partial \theta_j$ are
straightforward. With $r_{ij} = \lVert \bx_i - \bx_j \rVert$,
\[
\frac{\partial K_{ij}}{\partial \sigma_f^2}
= \frac{K_{ij}}{\sigma_f^2},
\qquad
\frac{\partial K_{ij}}{\partial \ell}
= K_{ij}\,\frac{r_{ij}^2}{\ell^3},
\qquad
\frac{\partial \Ky}{\partial \sigma_n^2} = I.
\]

\section{Maximizing aquisition functions}
It is easy to see that when the objective function is not concave, acquisition functions are generally non-concave and they are often multi-modal. Therefore, it might seem that the main problem of maximizing a non-concave function does not become any simpler by using Bayesian optimziation.

The crucial advantage of the approach lies in the fact that for a continuously differentiable kernel function, one can find explicit bounds on the Lipschitz constant, which is crucial in many methods for global optimization. While the truen objective function is typically also Lipschitz continuous, bounds on the Lipschitz constant cannot be obtained. 

Consider a zero mean Gaussian process defined through the continuous covariance kernel $k(\cdot, \cdot)$ with Lipschitz constant $L_k$ on the compact set $\mathbf{X}$.  Then it is easy to show (see, e.g. \cite{lederer2019uniform}) that the posterior mean function $\nu_N(\cdot)$ and standard deviation $\sigma_N(\cdot)$ of a Gaussian process conditioned on the training data $\left\{\left(\boldsymbol{x}_i, y_i\right)\right\}_{i=1}^N$ are continuous with Lipschitz constant $L_{\nu_N}$ and modulus of continuity $\omega_{\sigma_N}(\cdot)$ on $\mathbb{X}$ such that
$$
\begin{aligned}
L_{\nu_N} & \leq L_k \sqrt{N}\left\|\left(\boldsymbol{K}\left(\boldsymbol{X}_N, \boldsymbol{X}_N\right)+\sigma_n^2 \boldsymbol{I}_N\right)^{-1} \boldsymbol{y}_N\right\| \\
\omega_{\sigma_N}(\tau) & \leq \sqrt{2 \tau L_k\left(1+N\left\|\left(\boldsymbol{K}\left(\boldsymbol{X}_N, \boldsymbol{X}_N\right)+\sigma_n^2 \boldsymbol{I}_N\right)^{-1}\right\| \max _{\boldsymbol{x}, \boldsymbol{x}^{\prime} \in \mathbb{X}} k\left(\boldsymbol{x}, \boldsymbol{x}^{\prime}\right)\right)}
\end{aligned}
$$

Let $\operatorname{Lip}(k)=\left\{f: \mathcal{X} \rightarrow \mathbb{R}\right.$ s.t. $\left|f(x)-f\left(x^{\prime}\right)\right| \leq \left.k \cdot\left\|x-x^{\prime}\right\|_2, \forall\left(x, x^{\prime}\right) \in \mathcal{X}^2\right\}$ denote the class of $k$ Lipschitz functions defined on $\mathcal{X}$.
There are various global methods for this problem (see e.g. \cite{schweidtmann2021deterministic} or \cite{cortild2024global}). The intuition of why the knowledge for a Lipschitz bound is crucial can be obtained from the simplest stochastic method.
If one considers pure uniform sampling \cite{malherbe2017global} derive the following bounds.
 For any bounded and convex set $\mathcal{X} \subset \mathbb{R}^d$, we define its inner-radius as $\operatorname{rad}(\mathcal{X})=\max \{r> 0: \exists x \in \mathcal{X}$ such that $\mbox{Ball}(x, r) \subseteq \mathcal{X}\}$, its diameter as $\operatorname{diam}(\mathcal{X})=\max _{\left(x, x^{\prime}\right) \in \mathcal{X}^2}\left\|x-x^{\prime}\right\|_2$ and we denote by $\mu(\mathcal{X})$ its Lebesgue measure. Then we have the following result.
\begin{lemma} For any $f \in \operatorname{Lip}(k), n \in \mathbb{N}^{\star}$ and $\delta \in(0,1)$, we have with probability at least $1-\delta$,
$$
\max _{x \in \mathcal{X}} f(x)-\max _{i=1 \ldots \mathrm{n}} f\left(X_i\right) \leq k \cdot \operatorname{diam}(\mathcal{X}) \cdot\left(\frac{\ln (1 / \delta)}{n}\right)^{\frac{1}{d}} .
$$
\end{lemma}
Unfortunately, it is easy to see that uniform sampling faces a severe curse of dimensionality.
In low dimensional problems it can be employed to obtain useful bounds.

In practice, restarts of Newton based optimziation routines are typically used to obtain
in a global optimum. If one wants to establish probabilistic bounds, this method is fine during the BO iterations. The method generally works well, even if in each iteration only a local maximum of the acquisition function is identified. 
However, this paper’s certification argument requires a genuinely global upper bound on PI over the domain,
In the final iteration, to prove that PI is very small everywhere in the range, one has to use global methods to maximize the acquisition function.

\end{appendices}

\newpage

\end{document}